\documentclass[12pt]{article}
\usepackage{amsmath}
\usepackage{graphicx,psfrag,epsf}
\usepackage{enumerate}
\usepackage{natbib}
\usepackage{url} 
\usepackage{bbm}
\usepackage{bm}
\usepackage{graphicx}
\usepackage{subfig}
\usepackage{caption}

\usepackage{float}
\usepackage[figuresright]{rotating}
\usepackage{mathtools}
\usepackage{amsmath}
\usepackage[english]{babel}
\usepackage{times}
\usepackage{latexsym}
\usepackage{amssymb}

\usepackage{amsthm}

\newcommand{\blind}{0}
\def \ci {\perp\!\!\!\perp}
\def\bSig\mathbf{\Sigma}

\begin{document}

\def\spacingset#1{\renewcommand{\baselinestretch}%
{#1}\small\normalsize} \spacingset{1}

\if0\blind
{
  \title{\bf On Joint Estimation of Gaussian Graphical Models for Spatial and Temporal Data}
  \author{Zhixiang Lin\\
	Program in Computational Biology and Bioinformatics, Yale University\\
	Department of Statistics, Stanford University\\
   	Tao Wang \\
    	Department of Biostatistics, School of Public Health, Yale University\\
 	Can Yang\\
    	Department of Mathematics, Hong Kong Baptist University \\
    	Hongyu Zhao\thanks{
	To whom correspondence should be addressed. Supported in part by the National Science Foundation grant DMS-1106738 and the National Institutes of Health grants R01 GM59507 and P01 CA154295.} \\
    	Department of Biostatistics, School of Public Health, Yale University}
  \maketitle
} \fi

\if1\blind
{
  \bigskip
  \bigskip
  \bigskip
  \begin{center}
    {\LARGE\bf On Joint Estimation of Gaussian Graphical Models for Spatial and Temporal Data}
\end{center}
  \medskip
} \fi

\bigskip
\begin{abstract}
In this paper, we first propose a Bayesian neighborhood selection method to estimate Gaussian Graphical Models (GGMs). We show the graph selection consistency of this method in the sense that the posterior probability of the true model converges to one. When there are multiple groups of data available, instead of estimating the networks independently for each group, joint estimation of the networks may utilize the shared information among groups and lead to improved estimation for each individual network. Our method is extended to jointly estimate GGMs in multiple groups of data with complex structures, including spatial data, temporal data and data with both spatial and temporal structures. Markov random field (MRF) models are used to efficiently incorporate the complex data structures. We develop and implement an efficient algorithm for statistical inference that enables parallel computing. Simulation studies suggest that our approach achieves better accuracy in network estimation compared with methods not incorporating spatial and temporal dependencies when there are shared structures among the networks, and that it performs comparably well otherwise. Finally, we illustrate our method using the human brain gene expression microarray dataset, where the expression levels of genes are measured in different brain regions across multiple time periods.
\end{abstract}

\noindent%
{\it Keywords:}  Spatial and Temporal Data; Gaussian Graphical Model; Neighborhood Selection; Bayesian Variable Selection; Markov Random Field.
\vfill

\newpage
\spacingset{1.45} 
\section{Introduction}
\label{sec:intro}

The analysis of biological networks, including protein-protein interaction networks (PPI), biological pathways, transcriptional regulatory networks and gene co-expression networks, has led to numerous advances in the understanding of the organization and functionality of biological systems \citetext{e.g., \citealt{kanehisa2000kegg, shen2002network, rual2005towards,  zhang2005general}}. The work presented in this paper was motivated from the analysis of the human brain gene expression microarray data, where the expression levels of genes were measured in numerous spatial loci, which represent different brain regions, during different time periods of brain development \citep{kang2011spatio}. Although these data offer rich information on the network information among genes, only naive methods have been used for network inference. For example, \citet{kang2011spatio} pooled all the data from different spatial regions and time periods to construct a single gene network. However, only a limited number of data points are available for a specific region and time period, making region- and time- specific inference challenging.

Our aim here is to develop sound statistical methods to characterize the changes in the networks across time periods and brain regions, as well as the common network edges that are shared. This is achieved through a joint modeling framework to infer individual graphs for each brain region in each time period, where the degrees of spatial and temporal similarity are learnt adaptively from the data. Our proposed joint modeling framework may better capture the edges that are shared among graphs, and also allow the graphs to differ across brain regions and time periods.

We represent the biological network with a graph $G=(V, E)$ consisting of vertices $V=\{1,...,p\}$ and edges $E \subset V \times V$. In this paper, we focus on conditionally independent graphs, where $(i,j) \in E$ if and only if node $i$ and node $j$ are not conditionally independent given all the other nodes. Gaussian graphical models (GGMs) have been proven among the best to infer conditionally independent graphs. In GGM, the $p$-dimensional $X=(X_1, \ldots, X_p)$ is assumed to follow a multivariate Gaussian distribution $\mathcal{N}(\mu, \Sigma)$. Denote $\Theta=\Sigma^{-1}$ the precision matrix. It can be shown that the conditional independence of $X_i$ and $X_j$ is equivalent to $\Theta_{ij} \neq 0$:
$X_{i} \ci X_{j} \mid X _{V\backslash\{i,j\}} \Longleftrightarrow \Theta_{ij}=0$. In GGM, estimating the conditional independence graph is equivalent to estimating the non-zero entries in $\Theta$. Various approaches have been proposed to estimate the graph \citep{meinshausen2006high,yuan2007model,friedman2008sparse,cai2011constrained,dobra2011bayesian,wang2012bayesian,orchard2013bayesian}. Among these methods, \citet{friedman2008sparse} developed a fast and simple algorithm, named the graphical lasso (glasso), using a coordinate descent procedure for the lasso. They considered optimizing the penalized likelihood, with $\ell_1$ penalty on the precision matrix. As extensions of glasso, several approaches have been proposed to jointly estimate GGMs in multiple groups of data. \citet{guo2011joint} expressed the elements of the precision matrix for each group as a product of binary common factors and group-specific values. They incorporated an $\ell_1$ penalty on the common factors, to encourage shared sparse structure, and another $\ell_1$ penalty on the group-specific values, to allow edges included in the shared structure to be set to zero for specific groups. \citet{danaher2014joint} extended glasso more directly by extending the $\ell_1$ penalty for each precision matrix with additional penalty functions that encourage shared structure. They proposed two possible choices of penalty functions: 1. Fused lasso penalty that penalizes the difference of the precision matrices, which encourages common values among the precision matrices; 2. Group lasso penalty. \citet{chun2014gene} proposed a class of non-convex penalties for more flexible joint sparsity constraints. As an alternative to the penalized methods, \citet{peterson2014bayesian} proposed a Bayesian approach. They formulated the model in the $G$-Wishart prior framework and modeled the similarity of multiple graphs through a Markov Random Field (MRF) prior. However, their approach is only applicable when the graph size is small ($\sim 20$) and the number of groups is also small ($\sim5$). 

In this paper, we formulate the model in a Bayesian variable selection framework \citep{george1993variable, george1997approaches}.  \citet{meinshausen2006high} proposed a neighborhood selection procedure for estimating GGMs, where the neighborhood of node $i$ was selected by regressing on all the other nodes. Intuitively, our approach is the Bayesian analog of the neighborhood selection procedure. Our framework is applicable to the estimation of both single graph and multiple graphs. For the joint estimation of multiple graphs, we incorporate the MRF model. Compared with \citet{peterson2014bayesian}, we use a different MRF model and a different inferential procedure. One advantage of our approach is that it can naturally model complex data structures, such as spatial data, temporal data and data with both spatial and temporal structures. Another advantage is the computational efficiency. For the estimation of a single graph with $100$ nodes (the typical size of biological pathways is around that range), the computational time on a laptop is $\sim 30$ seconds for $1,000$ iterations of Gibbs sampling, which is $\sim 3$-folds faster than Bayesian Graphical Lasso, which implements a highly efficient block Gibbs sampler and is among the fastest algorithms for estimating GGMs in the Bayesian framework \citep{wang2012bayesian}. For multiple graphs, the computational time increases roughly linear with the number of graphs. Our procedure also enables parallel computing and the computational time can be substantially reduced if multicore processors are available. For single graph estimation, we show the graph selection consistency of the proposed method in the sense that the posterior probability of the true model converges to one.

The rest of the paper is organized as follows. We introduce the Bayesian neighborhood selection procedure for single graph and the extension to multiple graphs in Section 2. Model selection is discussed in Section 3. The theoretical properties are presented in Section 4. The simulation results are demonstrated in Section 5 and the application to the human brain gene expression microarray dataset is presented in Section 6. We conclude the paper with a brief summary in Section 7.

\section{Statistical Model and Methods}
\subsection{The Bayesian Neighborhood Selection Procedure} \label{BNS}
We first consider estimating the graph structure when there is only one group of data. Consider the $p$-dimensional multivariate normal random variable $X \sim \mathcal{N}(\mu, \Sigma)$. We further assume that $X$ is centered and $X \sim \mathcal{N}(0, \Sigma)$. Let $\Theta=\Sigma^{-1}$ denote the precision matrix. Let the $n \times p$ matrix $\mathbf{X}=(\mathbf{X}_1, \ldots ,\mathbf{X}_p)$ contain $n$ independent observations of $X$. For $A \subseteq \{1, \ldots, p\}$, define $\mathbf{X}_A = (\mathbf{X}_j, j \in A)$. Let $\Gamma_i$ denote the subset of $\{1, \ldots, p\}$, excluding the $i$th entry only. For any square matrix $C$, let $C_{i\Gamma_i}$ denote the $i$th row, excluding the $i$th element in that row. Consider estimating the neighborhood of node $i$. It is well known that the following conditional distribution holds:
\begin{equation} \label{cd_dist}
\mathbf{X}_{i} \mid \mathbf{X}_{\Gamma_i} \sim \mathcal{N}(-\mathbf{X}_{\Gamma_i}\Theta_{i\Gamma_i}^{T}\Theta_{ii}^{-1}, \Theta_{ii}^{-1}\bm{I}),
\end{equation}
where $\bm{I}$ is the $n \times n$ identity matrix, $\Theta_{ii}$ is a scalar and finding the neighborhood of $X_i$ is equivalent to estimating the non-zero coefficients in the regression of $X_i$ on $X_{\Gamma_i}$. Let $\bm{\beta}$ and $\bm{\gamma}$ be matrices of dimension $p \times p$, where $\bm\beta_{i\Gamma_i}=-\Theta_{ii}^{-1}\Theta_{i\Gamma_i}$ and $\bm{\gamma}$ is the binary latent state matrix. The diagonal elements in $\bm{\beta}$ and $\bm{\gamma}$ are not assigned values. Conditioning on $\gamma_{ij}$, $\beta_{ij}$ is assumed to follow a normal mixture distribution \citep{george1993variable, george1997approaches}:
\begin{equation*}
\text{ }  \beta_{ij} \mid \gamma_{ij} \sim (1-\gamma_{ij})\mathcal{N}(0, \tau_{i0}^2) + \gamma_{ij}\mathcal{N}(0, \tau_{i1}^2), \text{ for } j \in \Gamma_i, \\
 \end{equation*}
where $\tau_{i0}/\tau_{i1}=\delta$ and $0<\delta<1$. The prior on $\gamma_{ij}$ is Bernoulli: \\
\begin{equation*}
p(\gamma_{ij}=1)=1-p(\gamma_{ij}=0)=q.
 \end{equation*}
$\delta$, $\tau_{i1}$ and $q$ are prefixed hyperparameters and are discussed in the Supplementary Materials.
The off-diagonal entries in $\bm{\gamma}$ represent the presence or absence of the edges, which is the goal of our inference. 

Let $\bm{\sigma}=(\sigma_1, \ldots ,\sigma_p)$, where $\sigma_i^2=\Theta_{ii}^{-1}$. The inverse gamma (IG) conjugate prior is assumed for $\sigma_i^2$:
\begin{equation*}
\sigma_i^2 \mid \bm{\gamma} \sim IG(\nu_i/2,\lambda \nu_i/2).
 \end{equation*}
In this paper, we assume that $\nu_i=0$ and the IG prior reduces to a flat prior \citep{li2010bayesian}. 

For each node, we perform the Bayesian procedure to select the neighbors of that node. The precision matrix $\Theta$ is symmetric. If we let $\bm{\beta}_{i\Gamma_i}=-\Theta_{i\Gamma_i}$ instead of $-\Theta_{ii}^{-1}\Theta_{i\Gamma_i}$, the symmetric constraint can be satisfied by forcing $\bm{\beta}$ to be symmetric. However, this will lead to substantial loss in computational efficiency since the elements in $\bm{\beta}$ have to be updated one at a time, instead of one row at a time. Our simulation results suggest that the performance of the two models is comparable, whether or not the constraint on $\bm{\beta}$ is assumed (data not shown). Therefore, we do not constrain $\bm{\beta}$ in our inference. The selected neighborhood should be symmetric (the support of $\Theta$ is symmetric). \citet{meinshausen2006high} suggested using an or/and rule after their neighborhood selection procedure for each node. In our Bayesian procedure, the symmetric constrain can be incorporated naturally when sampling $\bm{\gamma}$ by setting $\gamma_{ij}=\gamma_{ji}$ for $j \neq i$. When there is no constraint assumed, the Bayesian procedure can be performed independently for each node. 

\subsection{Extension to mutiple graphs}
When there is similarity shared among multiple graphs, jointly estimating multiple graphs can improve inference. We propose to jointly estimate multiple graphs by specifying a Markov Random Field (MRF) prior on the latent states. Our model can naturally incorporate complex data structures, such as spatial data, temporal data and data with both spatial and temporal structures. Consider jointly estimating multiple graphs for data with both spatial and temporal structures. Denote $B$ the set of spatial loci and $T$ the set of time points. 
Our proposed model can be naturally implemented when there is missing data, i.e. no data points taken in certain locus at certain time point. For now, we assume that there is no missing data. The latent states for the whole dataset are represented by a $|B| \times |T| \times p \times p$ array $\bm{\gamma}$, where $|$ $|$ denotes the cardinality of a set. Let $\bm{\gamma}_{bt \cdot \cdot}$ denote the latent state matrix for locus $b$ at time $t$. In the real data example, $b$ is a categorical variable representing the brain region and $t$ is a discrete variable that represents the time period during brain development. Same as that in Section \ref{BNS}, the diagonal entries in $\bm{\gamma}_{bt \cdot \cdot}$ are not assigned values. 

Consider estimating the neighborhood of node $i$. Let $\gamma_{btij}$ denote the latent state for node $j\in\Gamma_i$ in locus $b$ at time $t$.  
Denote $\bm{\gamma_{\cdot \cdot ij}}=\{\gamma_{btij}: \forall b\in B, \forall t \in T\}$, $E^{s}_{ij}=\{(\gamma_{ijbt},\gamma_{ijb't'}): b\neq b' \text{, } t=t'\}$ and $E^{t}_{ij}=\{(\gamma_{btij},\gamma_{b't'ij}): b=b' \text{ and } |t-t'|=1 \}$. Here $E^{s}_{ij}$ contain all the pairs capturing spatial similarity and $E^{t}_{ij}$ contain all the pairs capturing temporal dependency between adjacent time periods. We do not consider the direction of the pairs: $(\gamma_{ijbt},\gamma_{ijb't'})$ and $(\gamma_{ijb't'},\gamma_{ijbt})$ are the same. Let $I_1(\cdot)$ and $I_0(\cdot)$ represent the indicator functions of $1$ and $0$, respectively.
The prior for $\bm{\gamma_{\cdot \cdot ij}}$ is specified by a pairwise interaction MRF model \citep{besag1986statistical, lin2015markov}:
\begin{equation}
\begin{split} \label{mrf}
p(\bm{\gamma}_{\cdot \cdot ij} \mid \bm{\Phi} ) \propto \exp & \Bigg{ \{ }
\eta_1 \sum_{b\in B, t \in T} I_{1}(\gamma_{ijbt})+ \\
& \eta_{s}\sum_{E^{s}_{ij}} \Big{[} I_0(\gamma_{btij})I_0(\gamma_{b't'ij})+I_1(\gamma_{btij})I_1(\gamma_{b't'ij}) \Big{ ] } + \\
& \eta_{t}\sum_{E^{t}_{ij}} \Big{[} I_0(\gamma_{btij})I_0(\gamma_{b't'ij})+I_1(\gamma_{btij})I_1(\gamma_{b't'ij}) \Big{]} \Bigg{ \} },
\end{split}
\end{equation}
and conditional independence is assumed:
\begin{equation}
p(\bm{\gamma} \mid \bm{\Phi} ) = \prod_{i}\prod_{j \in \Gamma_i}p(\bm{\gamma}_{\cdot \cdot ij} \mid \bm{\Phi} ), 
\end{equation}
where $\bm{\Phi}=\{ \eta_1, \eta_s, \eta_t\}$ are set to be the same for all $i$ and $j$. $\eta_1 \in \mathbb{R}$ and when there is no interaction terms, $1/(1+\exp(-\eta_1))$ corresponds to $q$ in the Bernoulli prior. $\eta_s \in\mathbb{R}$ represents the magnitude of spatial similarity and $\eta_t \in \mathbb{R}$ represents the magnitude of temporal similarity. In the simulation and real data example, $\eta_1$ is prefixed, whereas $\eta_s$ and $\eta_t$ are estimated from the dataset. Discussion on the choice of $\eta_1$ is provided in the Supplementary Materials. The priors on $\eta_s$ and $\eta_t$ are assumed to follow uniform distribution in $[0, 2]$.

Let $\bm{\gamma}_{\cdot \cdot ij}/\gamma_{btij}$ denote the subset of $\bm{\gamma}_{\cdot \cdot ij}$ excluding $\gamma_{btij}$. Then we have:
\begin{align}
p(\gamma_{btij} \mid \bm{\gamma}_{\cdot \cdot ij}/\gamma_{btij}, \bm{\Phi} ) = \frac{\exp\{\gamma_{btij} F(\gamma_{btij},\bm{\Phi})\}}{1+\exp\{F(\gamma_{btij},\bm{\Phi})\}},
 \label{cp}
\end{align}
where
\begin{align*}
F(\gamma_{btij},\bm{\Phi})=\eta_1 &+ \eta_{s}\sum_{b' \in B, b'\neq b}(2\gamma_{b'tij}-1) \\
&+\eta_{t}\{I_{t \neq 1}[2\gamma_{b(t-1)ij}-1]+I_{t \neq T}[2\gamma_{b(t+1)ij}-1]\}.
\end{align*}

In (\ref{mrf}), we made the following assumptions: a) the groups with different spatial labels are parallel to each other and they have the same magnitude of similarity and b) the time periods are evenly spaced and can be represented by integer labels. The first assumption can be relaxed by adjusting (\ref{mrf}) in two ways: 1. let $\eta_s$ vary for different pairs of loci (e.g. let $\eta_s$ be some parametric function of the pairwise distance); 2. adjust $E^{s}_{ij}$ to incorporate complex structure for the spatial loci (e.g. sub-groups or some graph to describe the adjacency of spatial loci). For the second assumption, $\eta_t$ can be adjusted to a parametric function of the time interval. When there is only spatial or only temporal structure in the dataset, model (\ref{mrf}) can be adjusted by removing the summation over the corresponding pairs. 

\section{Model selection}
For single graph and multiple graphs, the posterior probabilities are sensitive to the choices of hyperparameters. The ROCs are much more robust to the prior specification (Supplementary Materials). The posterior probability can be used as a score to rank the edges. One way of doing model selection is to sort the marginal posterior probabilities and select the top $K$ edges, where $K$ may depend on the prior knowledge of how sparse the graphs should be. An alternative way is to first perform thresholding on the marginal posterior probabilities to get an estimate of the graph structure $\hat{\mathcal{G}}$, and the precision matrix $\hat{\Theta}$ can be obtained by a fast iterative algorithm \citep{hastie2009elements}. By varying the threshold, different $\hat{\Theta}$s are obtained and some model selection criteria (for example, BIC) can be incorporated to select a model. 
\section{Theoretical Properties}
We rewrite $p$ as $p_n$ to represent a sequence $p_n$ that changes with $n$. 
Let $1 \leq p^* \leq p_n$. Throughout, we assume that $\mathbf{X}$ satisfies the sparse Riesz condition \citep{zhang2008sparsity} with rank $p^*$; that is, there exist some constants $0 < c_1 < c_2 < \infty$ such that
$$c_1 \leq \frac{\| \mathbf{X}_A u\|^2}{n \|u\|^2} \leq c_2,$$
for any $A \subseteq \{1, \ldots, p_n\}$ with size $|A| = p^*$ and any nonzero vector $u \in \mathbb{R}^{p^*}$.

Consider estimating the neighborhood for the $i$th node. We borrow some notations from \citet{narisetty2014bayesian}. For the simplicity of notation, let $\bm \beta^i \equiv \bm\beta_{i\Gamma_i} = -\Theta_{i\Gamma_i}^{T}\Theta_{ii}^{-1}$ and $\bm \gamma^i \equiv \bm \gamma_{i\Gamma_i}$. Write $\tau_{i0}$, $\tau_{i1}$ and $q$ as $\tau_{0n}$, $\tau_{1n}$ and $q_n$, respectively, to represent sequences that change with $n$.  We use a $(p_n - 1) \times 1$ binary vector $k^i$ to index an arbitrary model. The corresponding design matrix and parameter vector are denoted by $\mathbf{X}_{k^i}\equiv (\mathbf{X}_{\Gamma_i})_{k^i}$ and $\bm\beta^i_{k^i}$, respectively. Let $t^i$ represent the true neighborhood of node $i$.

Denote by $\lambda_{\max}(\cdot)$ and $\lambda_{\min}(\cdot)$ the largest and smallest eigenvalues of a matrix, respectively. For $v > 0$, define
$$m(v) \equiv m_n(v) = (p_n - 1) \wedge \frac{n}{(2 + v) \log(p_n - 1)}$$ and $$\lambda_{m, i}(v) = \min_{k^i: |k^i|\leq m(v)} \lambda_{\min}\left(\frac{\mathbf{X}_{k^i}' \mathbf{X}_{k^i}}{n}\right).$$
For $K > 0$, let
$$\Delta_i(K) = \min_{\{k^i: |k^i| \leq K |t^i|, k^i \not\supset t^i\}}\|(I - P_{k^i})\mathbf{X}_{t^i}\bm\beta^i_{t^i}\|_2^2,$$ where $|k^i|$ denotes the size of the model $k^i$ and $P_{k^i}$ is the projection matrix onto the column space of $\mathbf{X}_{k^i}$.

For sequences $a_n$ and $b_n$, $a_n \sim b_n$ means $a_n / b_n \rightarrow c$ for some constant $c > 0$, $a_n \prec b_n$ (or $b_n \succ a_n$) means $a_n = o(b_n)$, and $a_n \preceq b_n$ (or $b_n \succeq a_n$) means $a_n = O(b_n)$. We need the following conditions.
\begin{enumerate}
\item[(A)] $p_n \rightarrow \infty$ and $p_n = O(n^\theta)$ for some $\theta > 0$;
\item[(B)] $q_n = p_n^{\alpha -1}$ for some $0 \leq \alpha < 1 \wedge (1 / \theta)$;
\item[(C)] $n\tau_{0n}^2 = o(1)$ and $n \tau_{1n}^2 \sim n \vee p_n^{2 + 2\delta_1}$ for some $\delta_1 > 1 + \alpha$;
\item[(D)] $|t^i| \prec n / \log p_n$ and $\|\bm\beta^i_{t^i}\|_2^2 \prec \tau_{1n}^2 \log p_n$;
\item[(E)] there exist $1 + \alpha < \delta_2 < \delta_1$ and $K > 1 + 8 / (\delta_2  - 1 - \alpha)$ such that, for some large $C > 0$, $\Delta_i(K)/ \sigma_i^2 > C |t^i| \log(n \vee p_n^{2 + 2\delta_1})$;
\item[(F)] $p^* \geq (K + 1) |t^i|$;
\item[(G)] $\lambda_{\max}(\mathbf{X}' \mathbf{X} / n) \prec (n\tau_{0n}^2)^{-1} \wedge (n\tau_{1n}^2)$ and there exist some $0 < v < \delta_2$ and $0 < \kappa < 2(K - 1)$ such that $$\lambda_{m, i}(v) \succeq \frac{(n \vee p_n^{2 + 2\delta_2})}{n \tau_{1n}^2} \vee p_n^{-\kappa}.$$
\end{enumerate}
{\sc Theorem 1.} Assume conditions (A)-(G). For some $c > 0$ and $s > 1$ we have, with probability at least $1 - c p_n^{-s}$, $P(\bm\gamma^i = t^i \mid \mathbf{X}, \sigma_i^2) > 1 - r_n$, where $r_n$ goes to 0 as the sample size increases to $\infty$.


To establish graph-selection consistency, we need slightly stronger conditions than (D)-(G). Let
$$t^* = \max_{1 \leq i \leq p_n} |t^i|,
\Delta^*(K) = \min_{1 \leq i \leq p_n} (\Delta_i(K)/ \sigma_i^2) \ {\rm and} \ \lambda^*_{m}(v) = \min_{1 \leq i \leq p_n} \lambda_{m, i}(v).$$
\begin{enumerate}
\item[(D')] $t^* \prec n / \log p_n$ and $\max_{1 \leq i \leq p_n} \|\bm\beta^i_{t^i}\|_2^2 \prec \tau_{1n}^2 \log p_n$;
\item[(E')] there exist $1 + \alpha < \delta_2 < \delta_1$ and $K > 1 + 8 / (\delta_2  - 1 - \alpha)$ such that, for some large $C > 0$, $\Delta^*(K) > C \log(n \vee p_n^{2 + 2\delta_1})$;
\item[(F')] $p^* \geq (K + 1) t^*$;    
\item[(G')] $\lambda_{\max}(\mathbf{X}' \mathbf{X} / n) \prec (n\tau_{0n}^2)^{-1} \wedge (n\tau_{1n}^2)$ and there exist some $0 < v < \delta_2$ and $0 < \kappa < 2(K - 1)$ such that $$\lambda^*_{m}(v) \succeq \frac{(n \vee p_n^{2 + 2\delta_2})}{n \tau_{1n}^2} \vee p_n^{-\kappa}.$$
\end{enumerate}
Let $\mathcal{G}$ denote the true graph structure and $\bm{\gamma}$ is the latent state matrix for all the nodes. \\
{\sc Theorem 2.} Assume conditions (A)-(C) and (D')-(G'). We have, as $n \rightarrow \infty$, $P( \bm{\gamma} = \mathcal{G} \mid \mathbf{X}, \bm{\sigma}^2) \rightarrow 1$.

\section{Simulation examples}
\subsection{Joint estimation of multiple graphs} \label{multiple}
We first considered the simulation of three graphs. For all three graphs, $p=100$ and $n=150$. We first simulated the graph structure. We randomly selected $5\%$ or $10\%$ among all the possible edges and set them to be edges in graph 1. For graphs 2 and 3, we removed a portion ($20\%$ or $100\%$) of edges that were present in graph 1 and added back the same number of edges that were not present in graph 1. $20\%$ represents the case that there is moderate shared structure. $100\%$ represents the extreme case that there is little shared structure other than those shared by chance. For the entries in the precision matrices, we considered two settings: a) the upper-diagonal entries were sampled from uniform $[-0.4, -0.1]\cup [0.1, 0.4]$ independently and then set the matrix to be symmetric b) Same as that in a), except that for the shared edges, the corresponding entries were set to be the same. To make the precision matrix positive definite, we set the diagonal entry in a row to be the sum of absolute values of all the other entries in that row, plus $0.5$. 

\begin{figure}[H]
\centering
\subfloat[Sparsity$\sim$0.05, change=0.2, different entry values][Sparsity$\sim$0.05, change=0.2, \\   \text{ } \text{ } \text{ } \text{ } \text{ } different entry values]{
\includegraphics[width=0.28\textwidth]{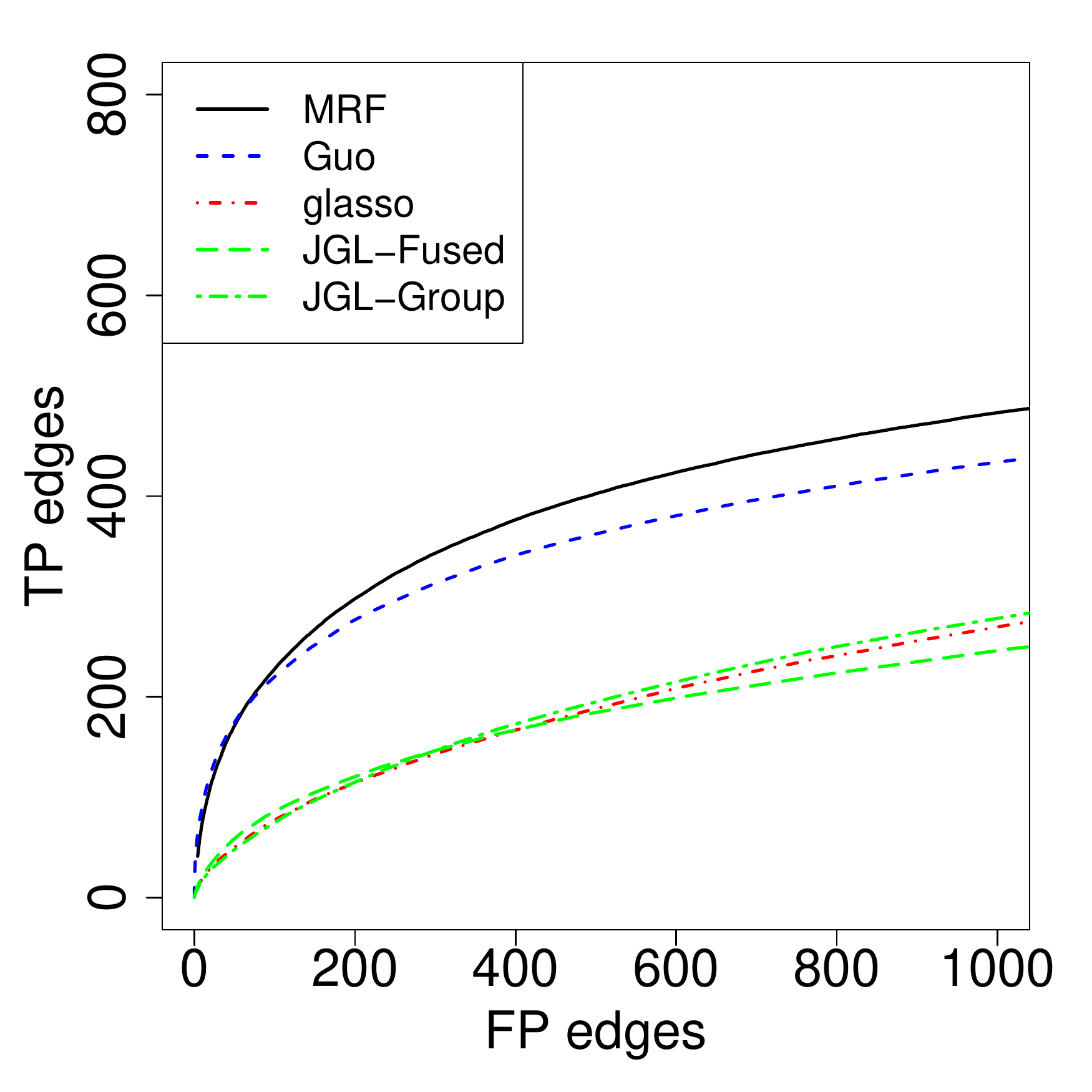}}
\subfloat[Sparsity$\sim$0.1, change=0.2, different entry values][Sparsity$\sim$0.1, change=0.2, \\ \text{ } \text{ } \text{ } \text{ } \text{ } different entry values]{
\includegraphics[width=0.28\textwidth]{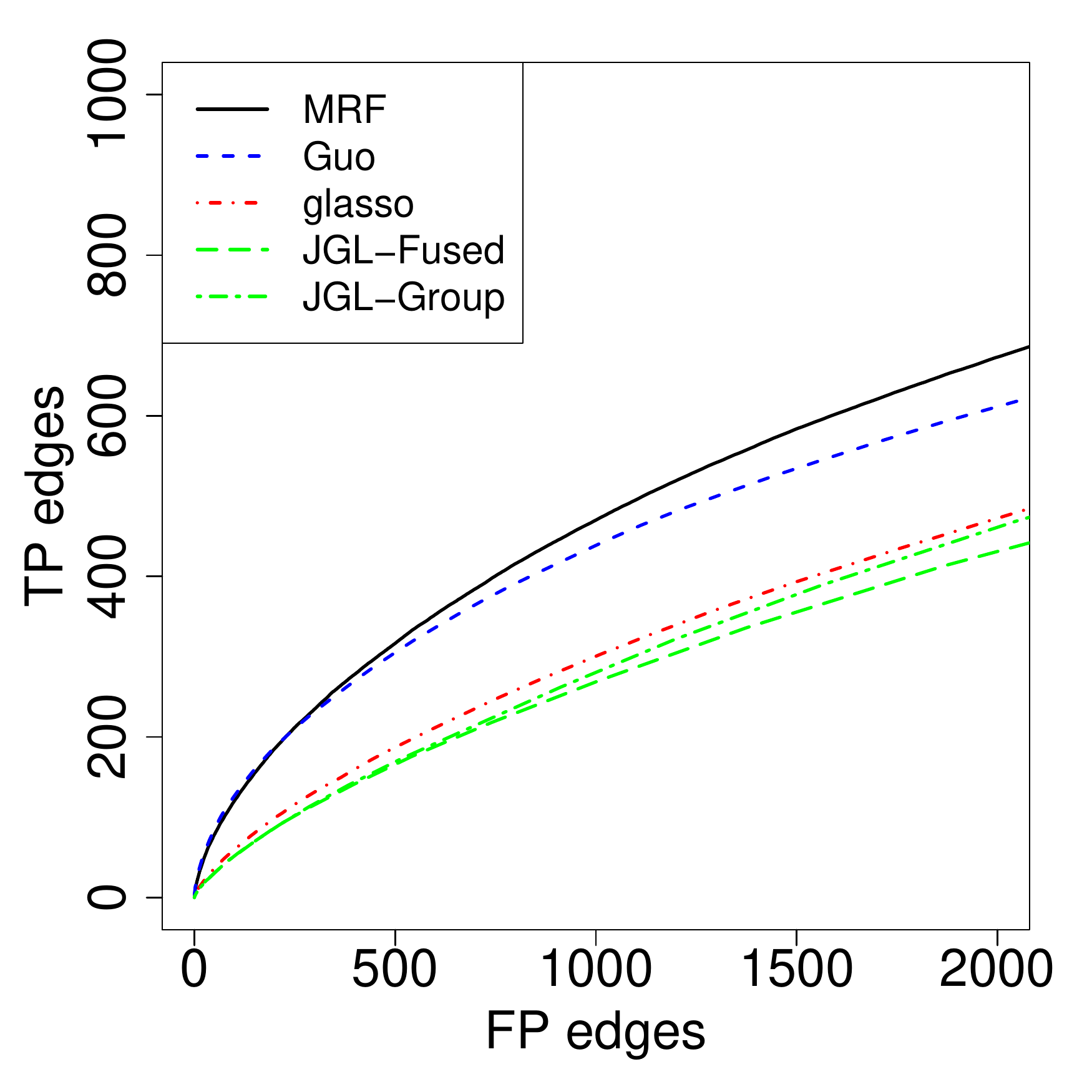}}
\subfloat[Sparsity$\sim$0.05, change=1, different entry values][Sparsity$\sim$0.05, change=1, \\ \text{ } \text{ } \text{ } \text{ } \text{ } different entry values]{
\includegraphics[width=0.28\textwidth]{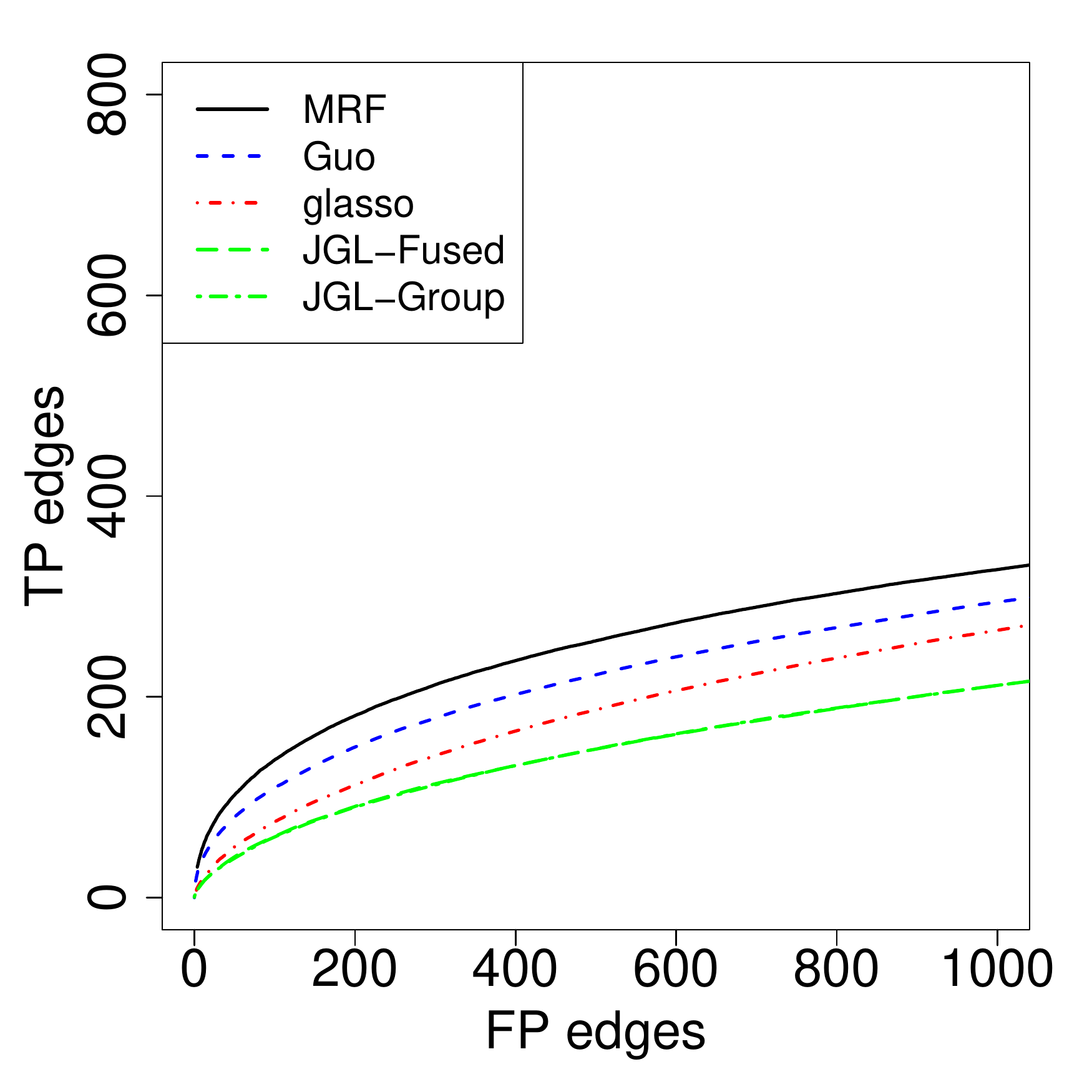}} \\
\subfloat[Sparsity$\sim$0.1, change=1, different entry values][Sparsity$\sim$0.1, change=1, \\ \text{ } \text{ } \text{ } \text{ } \text{ } different entry values]{
\includegraphics[width=0.28\textwidth]{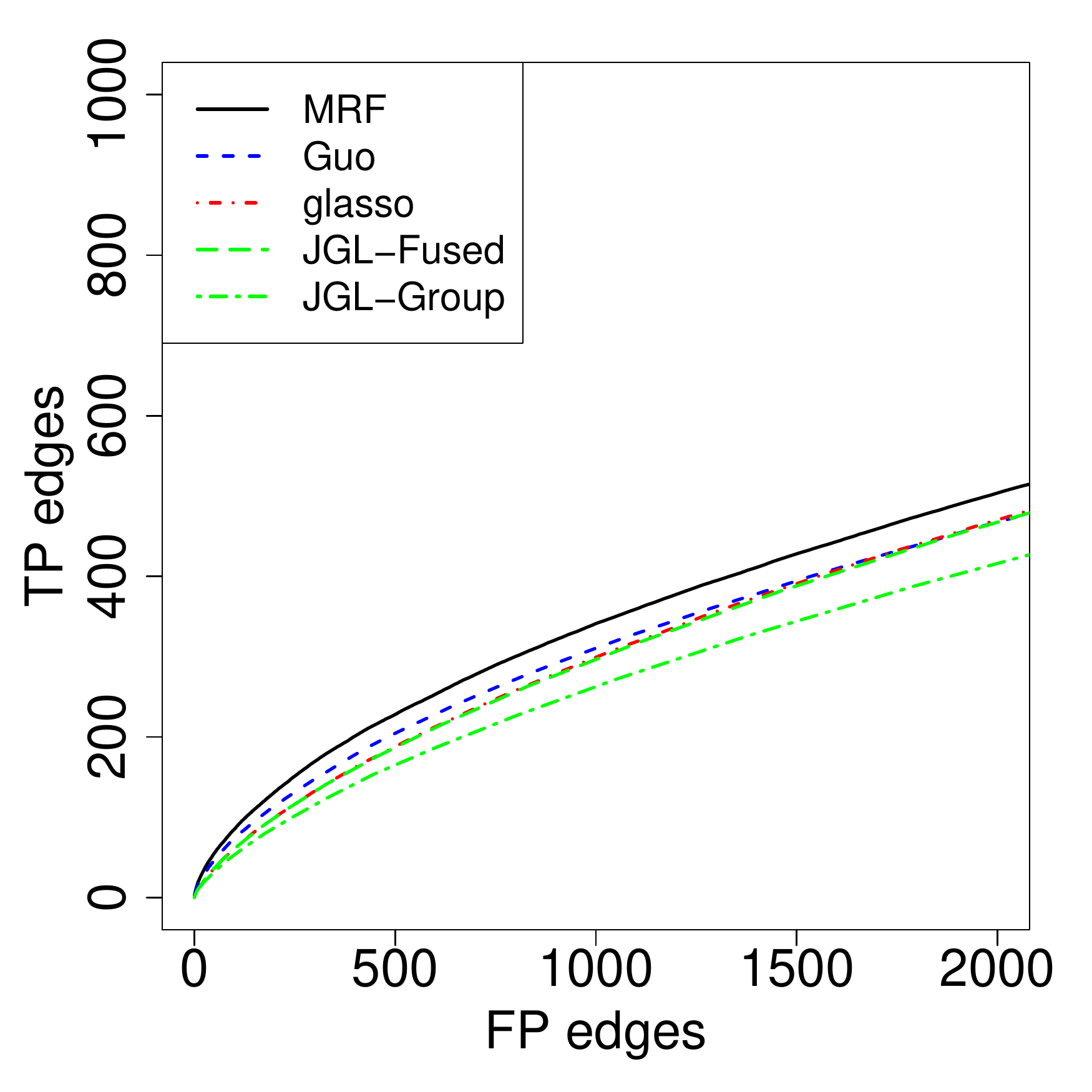}}
\subfloat[Sparsity$\sim$0.05, change=0.2, same entry values][Sparsity$\sim$0.05, change=0.2, \\ \text{ } \text{ } \text{ } \text{ } \text{ } same entry values]{
\includegraphics[width=0.28\textwidth]{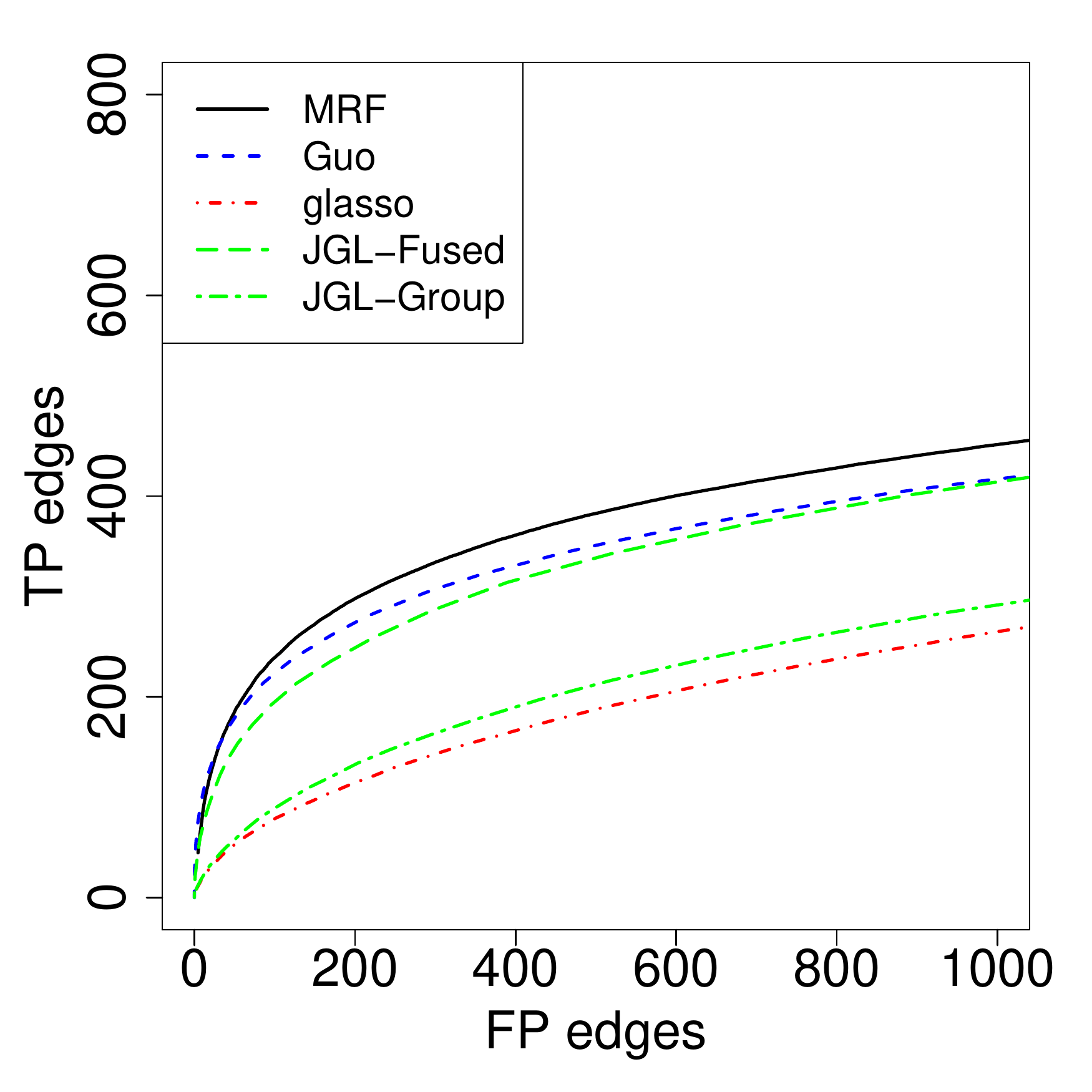}}
\subfloat[Sparsity$\sim$0., change=0.2, same entry values][Sparsity$\sim$0., change=0.2, \\ \text{ } \text{ } \text{ } \text{ } \text{ } same entry values]{
\includegraphics[width=0.28\textwidth]{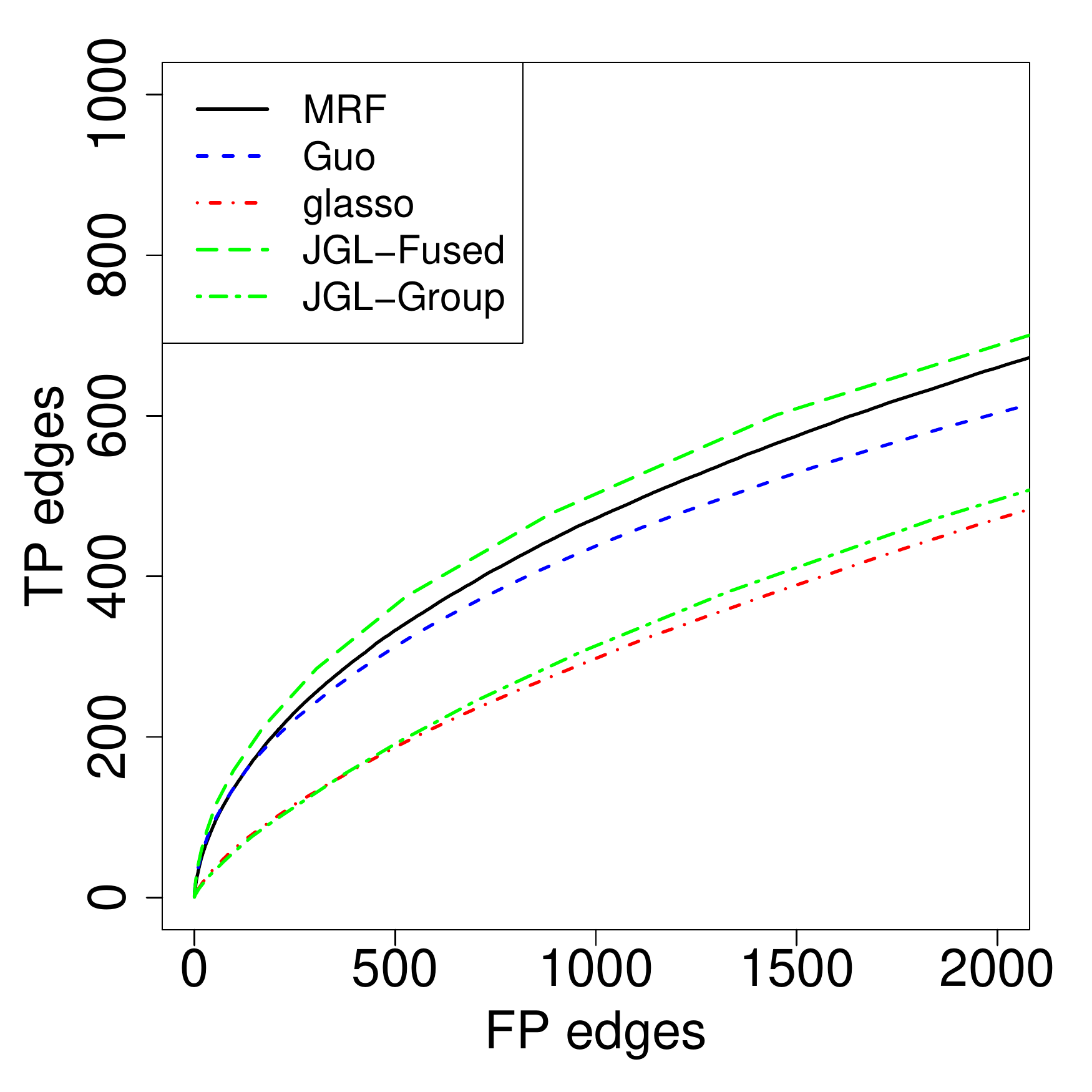}} \\
\subfloat[Sparsity$\sim$0.05, change=1, same entry values][Sparsity$\sim$0.05, change=1, \\ \text{ } \text{ } \text{ } \text{ } \text{ } same entry values]{
\includegraphics[width=0.28\textwidth]{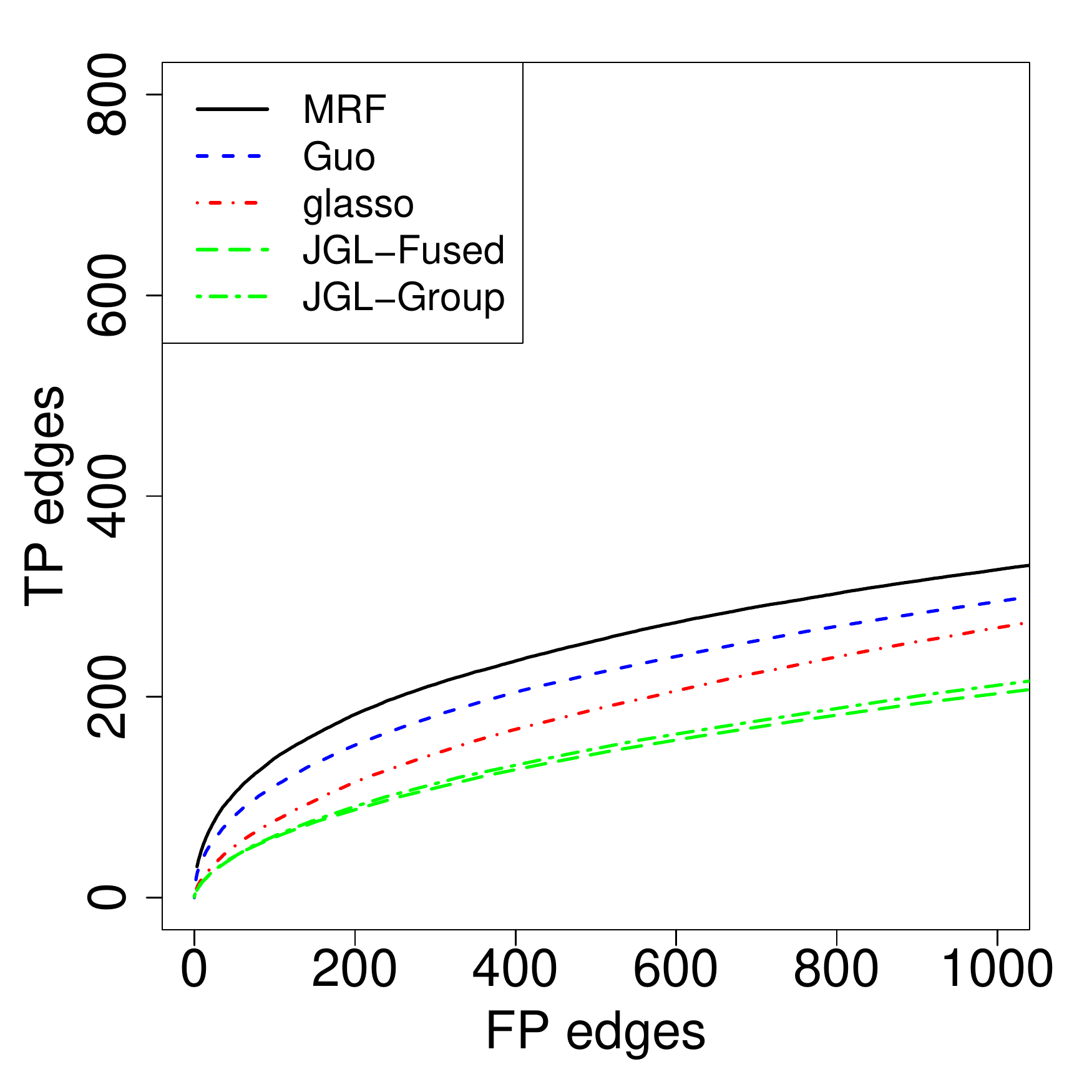}}
\subfloat[Sparsity$\sim$0.1, change=1, same entry values][Sparsity$\sim$0.1, change=1, \\ \text{ } \text{ } \text{ } \text{ } \text{ } same entry values]{
\includegraphics[width=0.28\textwidth]{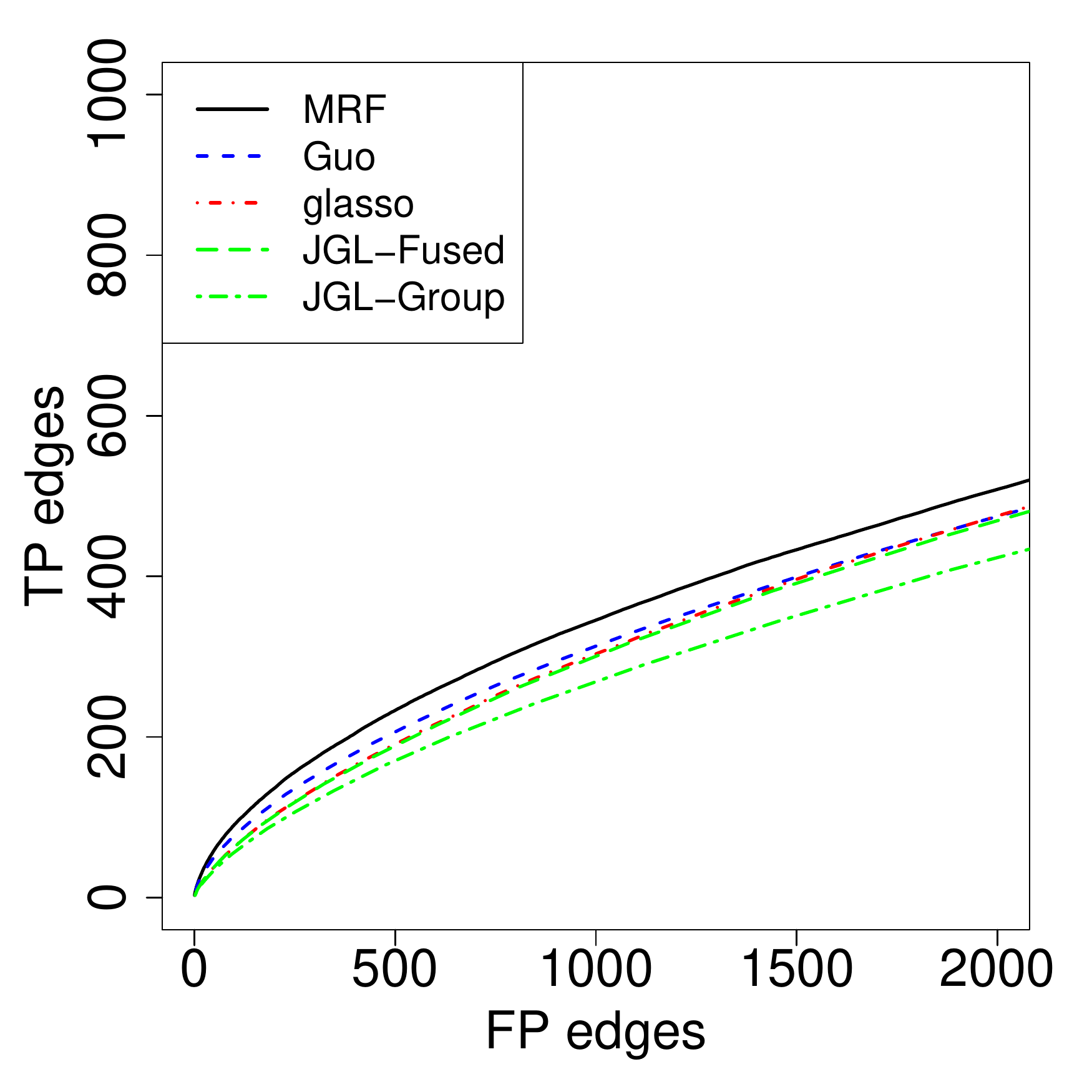}}
\caption{Comparisons of different models for the estimation of three graphs. For the shared edges, the corresponding entries in the precision matrices take the same (``same entry values'') or different (``different entry values'') non-zero values. The x-axis was truncated to be slightly larger than the total number of true positive edges. The curves represent the average of 100 independent runs.}
\label{graph3}
\end{figure}

The simulation results are presented in Figure \ref{graph3}. Our method (MRF) was compared with Guo's method \citep{guo2011joint}, JGL \citep{danaher2014joint} and graphical lasso (\textit{glasso}) \citep{friedman2008sparse}. In \textit{glasso}, the graphs are estimated independently. In JGL, there are two options, fused lasso (JGL-Fused) and group lasso (JGL-Group). For Guo's method, \textit{glasso} and JGL, we varied the sparsity parameter to generate the curves. For our method, we varied the threshold for the marginal posterior probabilities of $\bm{\gamma}$ to generate the curves. There are two tuning parameters in JGL, $\lambda_1$ and $\lambda_2$, where $\lambda_1$ controls sparsity and $\lambda_2$ controls the strength of sharing. We performed a grid search for $\lambda_2$ in $\{0, 0.05,...,0.5\}$ and selected the best curve. In Figure \ref{graph3}, our method performed slightly better than Guo's method. When there is little shared structure among graphs, our method performed slightly better than \textit{glasso}, which is possibly due to the fact that we used a different modeling framework. When the entries were different for the shared edges, JGL-Fused did not perform well. However, when the entries were the same, JGL-Fused performed much better. The fused lasso penalty encourages entries in the precision matrix to be the same and JGL-Fused gains efficiency when the assumption is satisfied. 

\subsection{Joint estimation of multiple graphs with temporal dependency} \label{hmm}
\begin{figure} 
\centering
\subfloat[$n=100$, $|T|=10$]{
\includegraphics[width=0.32\textwidth]{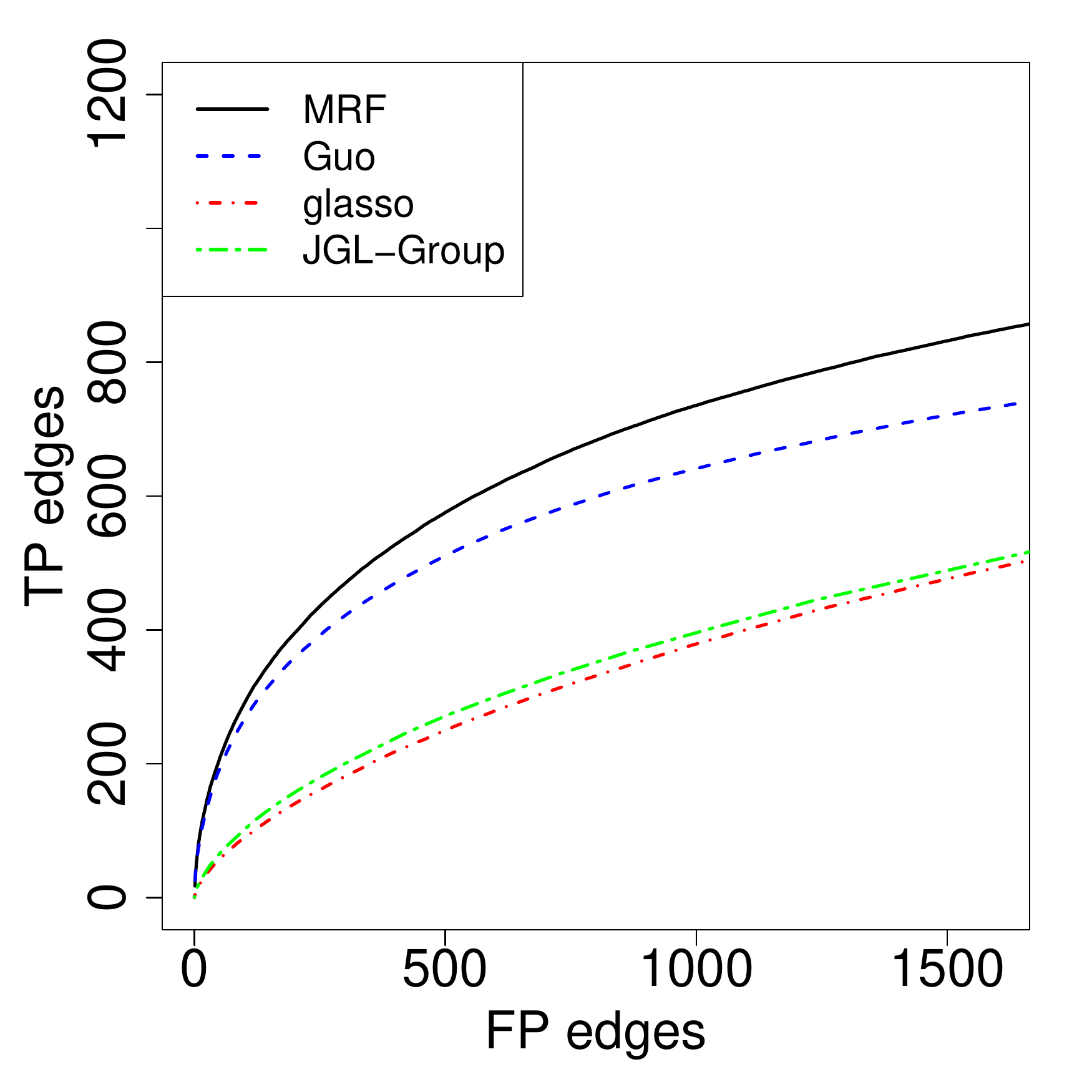}}
\subfloat[$n=200$, $|T|=10$]{
\includegraphics[width=0.32\textwidth]{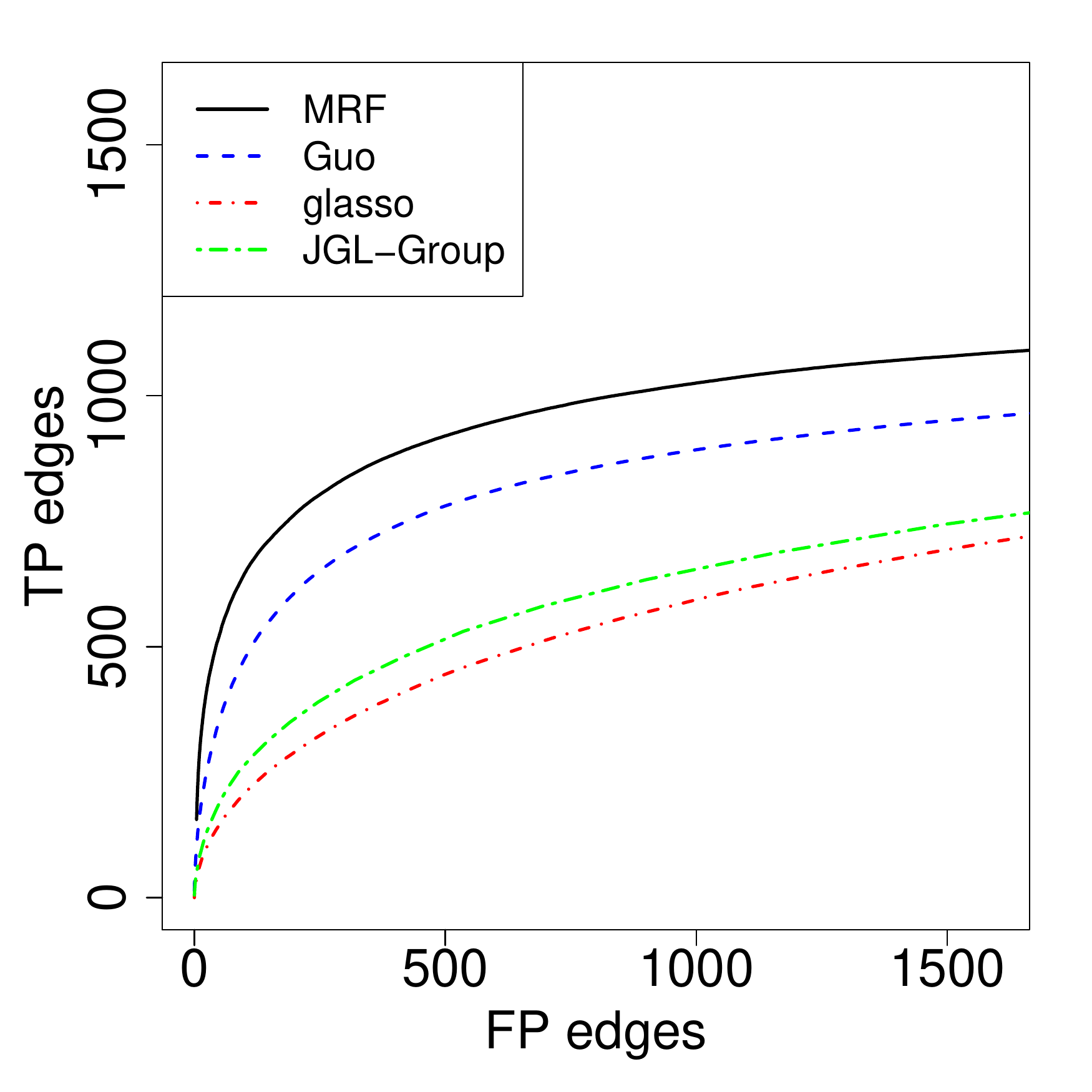}}
\subfloat[$n=100$, $|T|=30$]{
\includegraphics[width=0.32\textwidth]{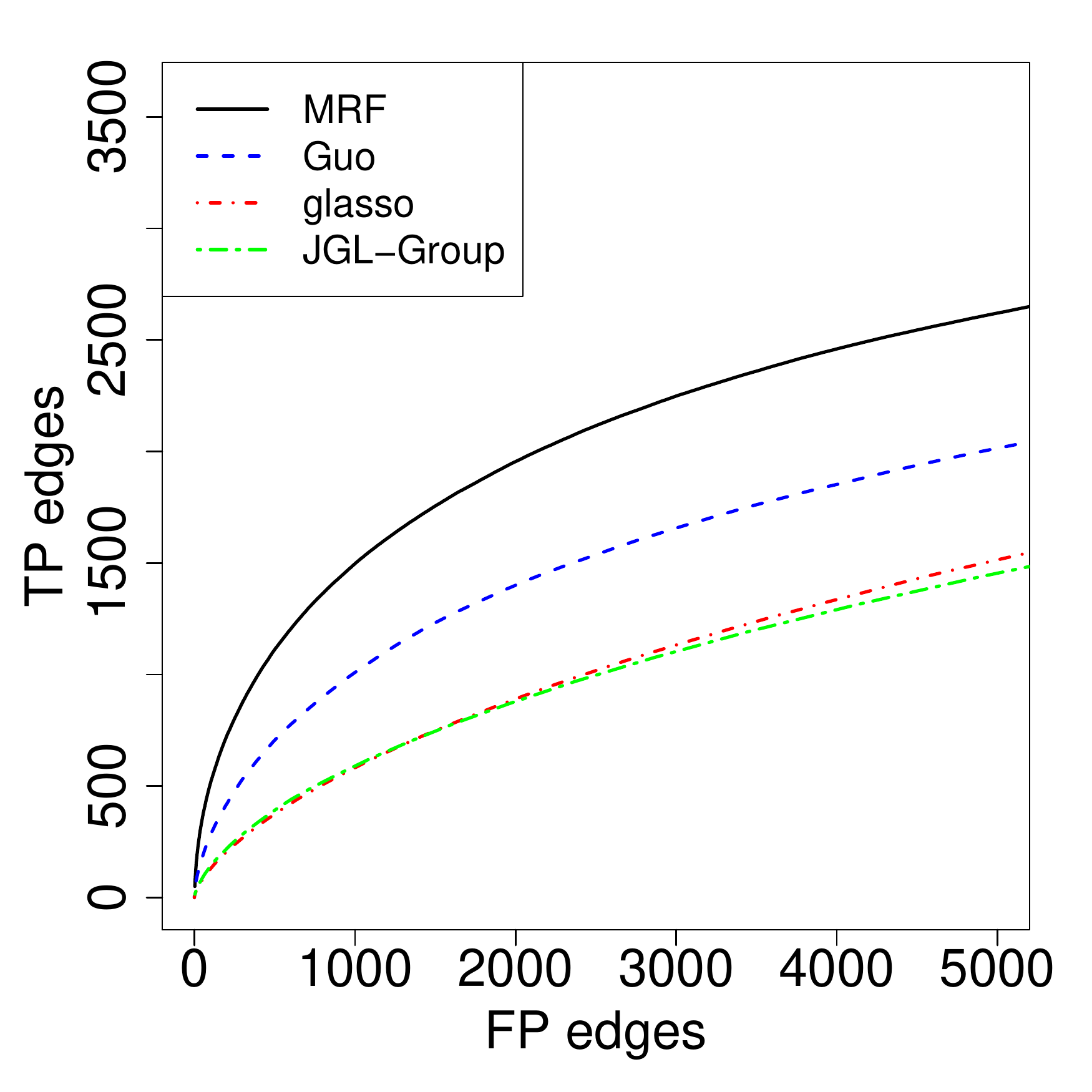}}
\caption{Comparisons of different models for the estimation of mutiple graphs with temporal dependency. The x-axis was truncated to be slightly larger than the total number of true positive edges. The curves represent the average of 100 independent runs.}
\label{temporal}
\end{figure}

In this setting, we assumed that the graph structure evolved over time by Hidden Markov Model (HMM). We set $p=50$. At time $t=1$, we randomly selected $10\%$ among all the possible edges and set them to be edges. At time $t+1$, we removed $20\%$ of the edges at time $t$ and added back the same number of edges that were not present at time $t$. The entries in the precision matrix were set the same as that in a) in Section \ref{multiple}. We present the simulation results in Figure \ref{temporal}, varying
$n$ and $|T|$. We compared our method with Guo's and JGL-Group, where the graphs were treated as parallel. Our method performed better than Guo's method and JGL-Group in all three settings, and the difference was greater when either $n$ or $|T|$ increases. We did not include JGL-Fused in the comparison as the computational time for JGL-Fused increases substantially when the number of graphs is more than a few.

\subsection{Joint estimation of multiple graphs with both spatial and temporal dependency}
\begin{figure} 
\centering
\subfloat[perturbation=0.1]{
\includegraphics[width=0.32\textwidth]{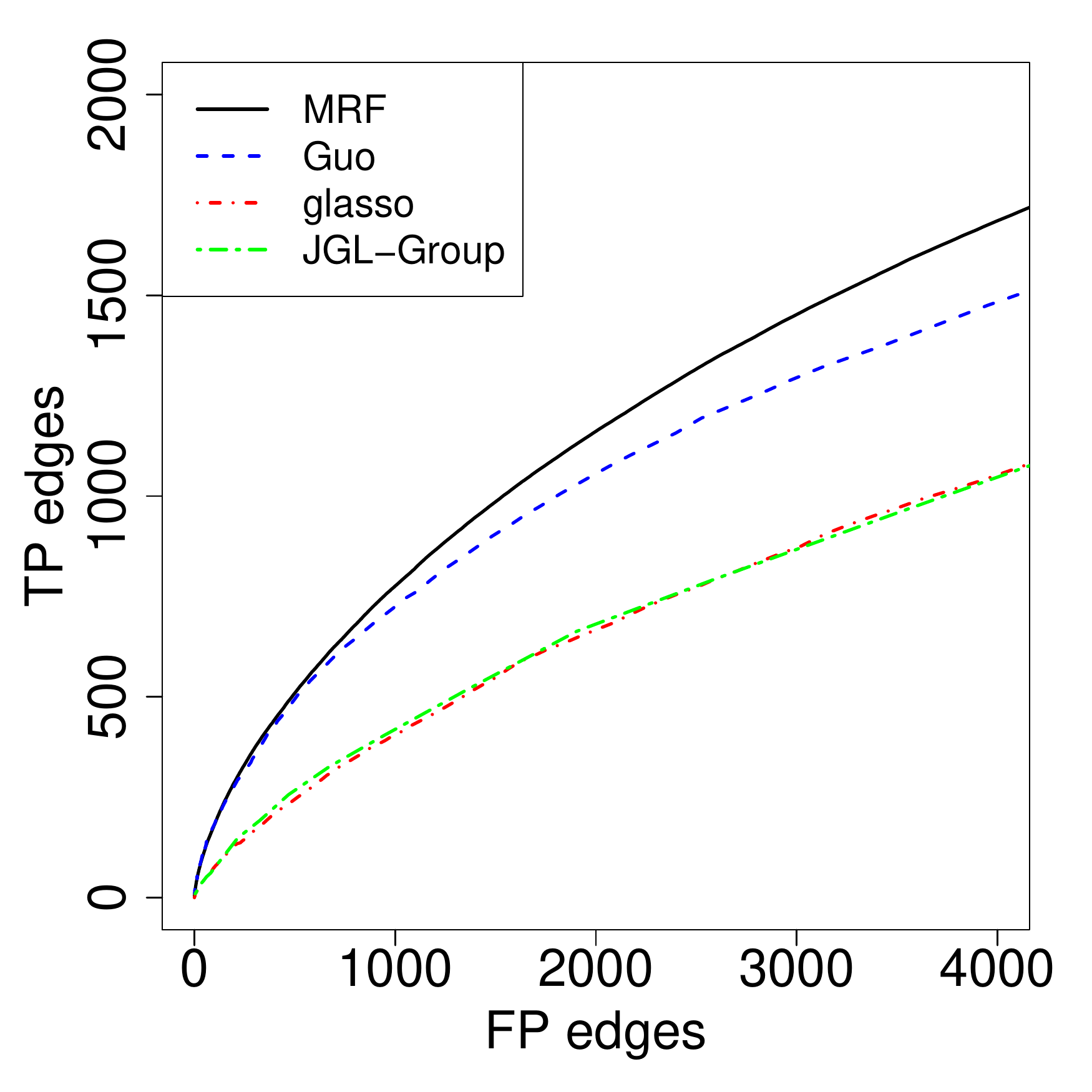}}
\subfloat[perturbation=0.2]{
\includegraphics[width=0.32\textwidth]{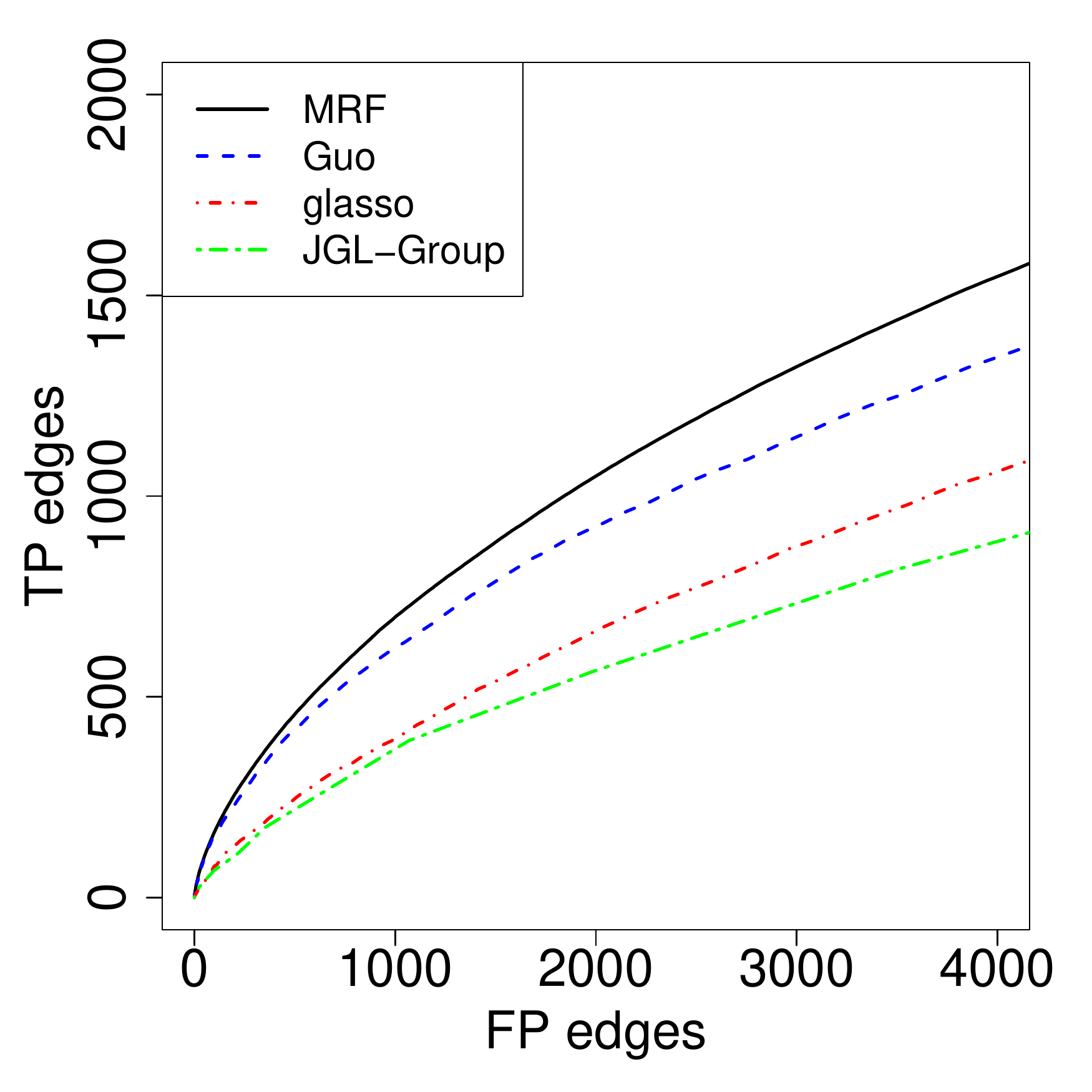}}
\subfloat[perturbation=0.5]{
\includegraphics[width=0.32\textwidth]{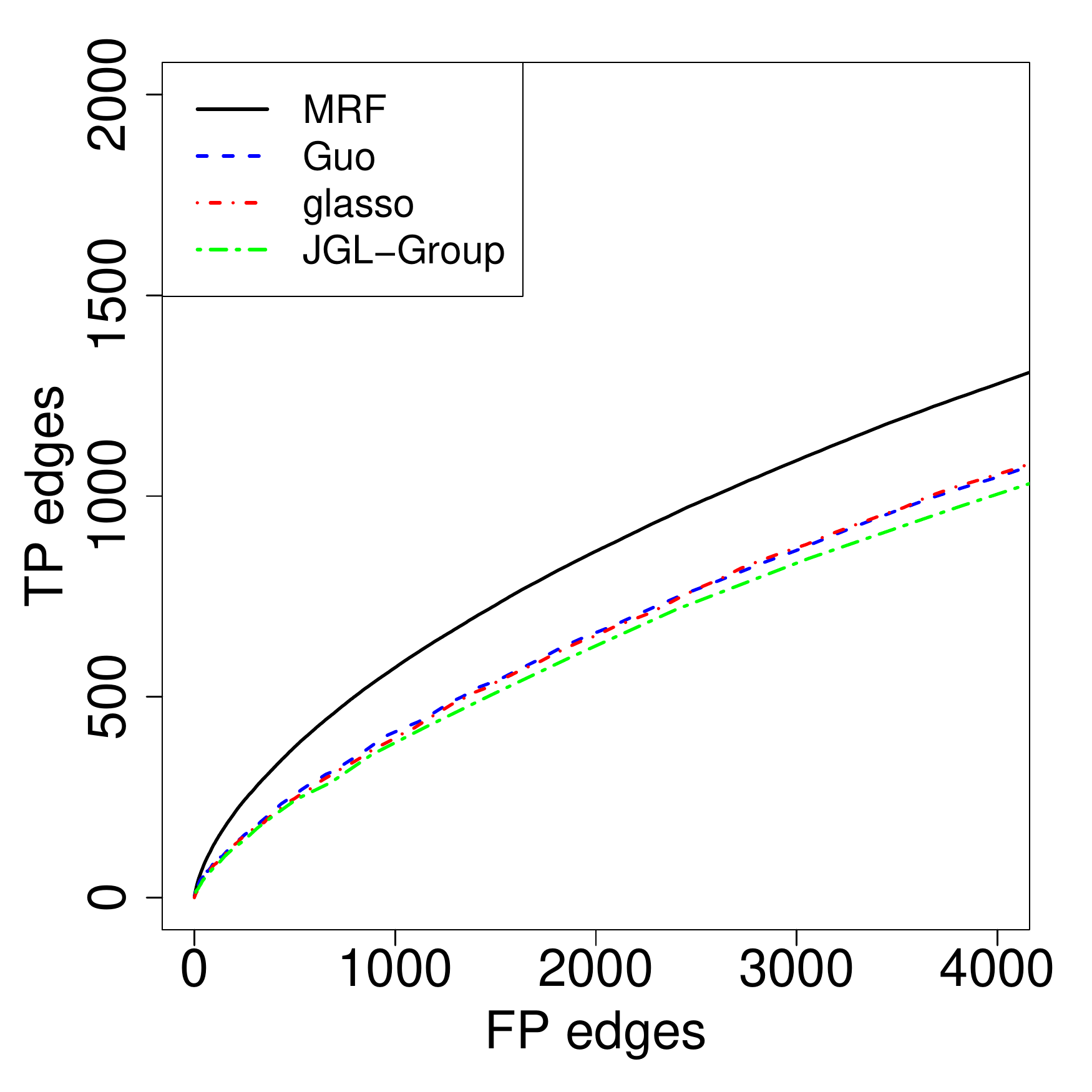}}
\caption{Comparisons of different models for the estimation of mutiple graphs with temporal and spatial dependency. The x-axis was truncated to be slightly larger than the total number of true positive edges.The curves represent the average of 100 independent runs.}
\label{spatialtemporal}
\end{figure}

We simulated graphs in $|B|=3$ spatial loci and $|T|=10$ time periods. We set $p=50$, $n=100$, and sparsity$\sim 0.1$. We first set the graphs in different loci at the same time point to be the same. The graph structure evolved over time by HMM similarly as that in Section \ref{hmm}, and $40\%$ of the edges changed between adjacent time points. For all graphs, we then added some perturbations by removing a portion ($10\%$, $20\%$, $50\%$) of edges and adding back the same number of edges. For simplicity, we treated the spatial loci as parallel and did not simulate more complex structures. The entries in the precision matrix were set the same as that in a) in Section \ref{multiple}. The simulation results are presented in Figure \ref{spatialtemporal}. Our method achieved better performance than all the other methods.

\subsection{Computational time}
\begin{figure}
\centering
\subfloat[Single Graph]{
\label{node}
\includegraphics[width=0.35\textwidth]{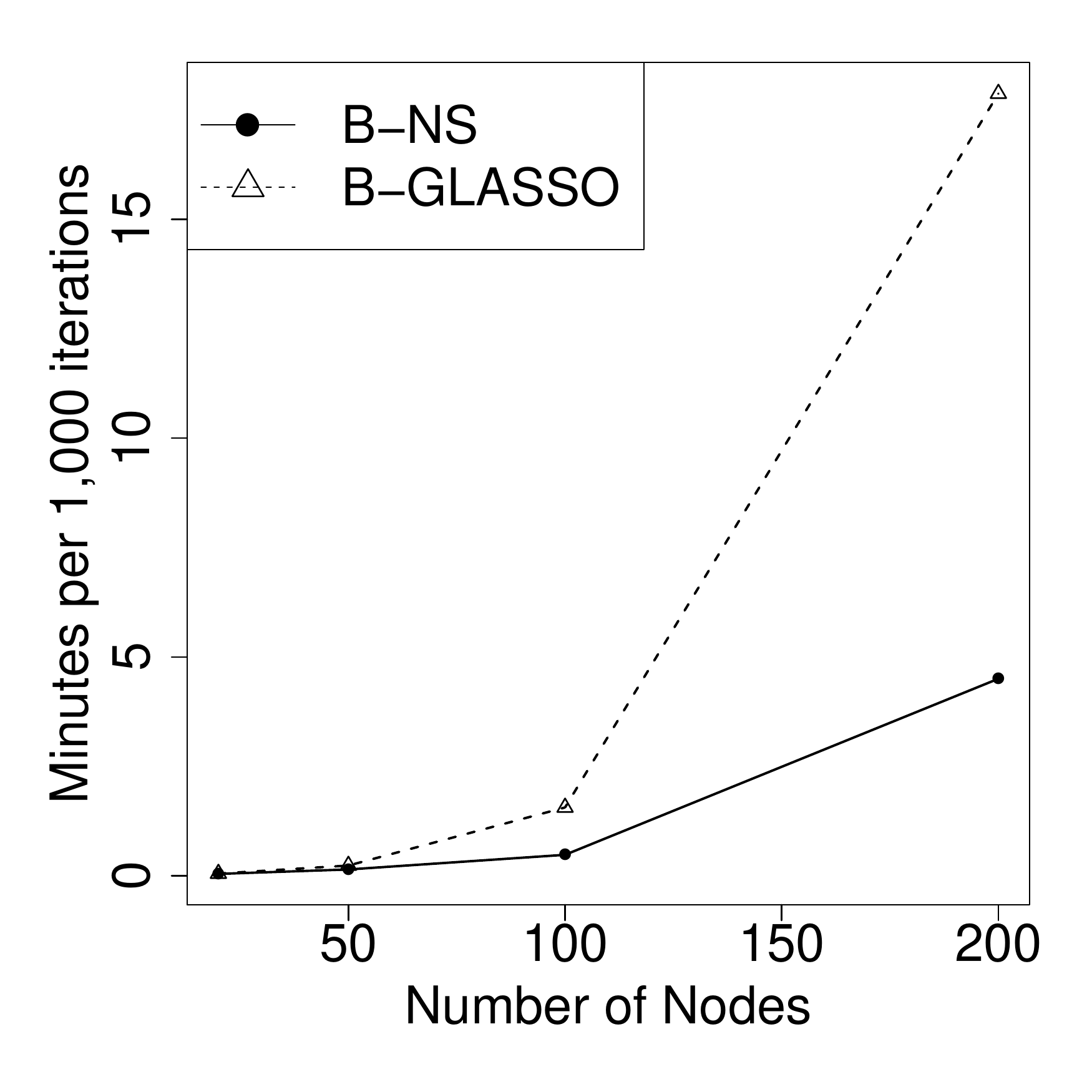}}
\subfloat[Multiple Graphs]{
\label{nsam}
\includegraphics[width=0.35\textwidth]{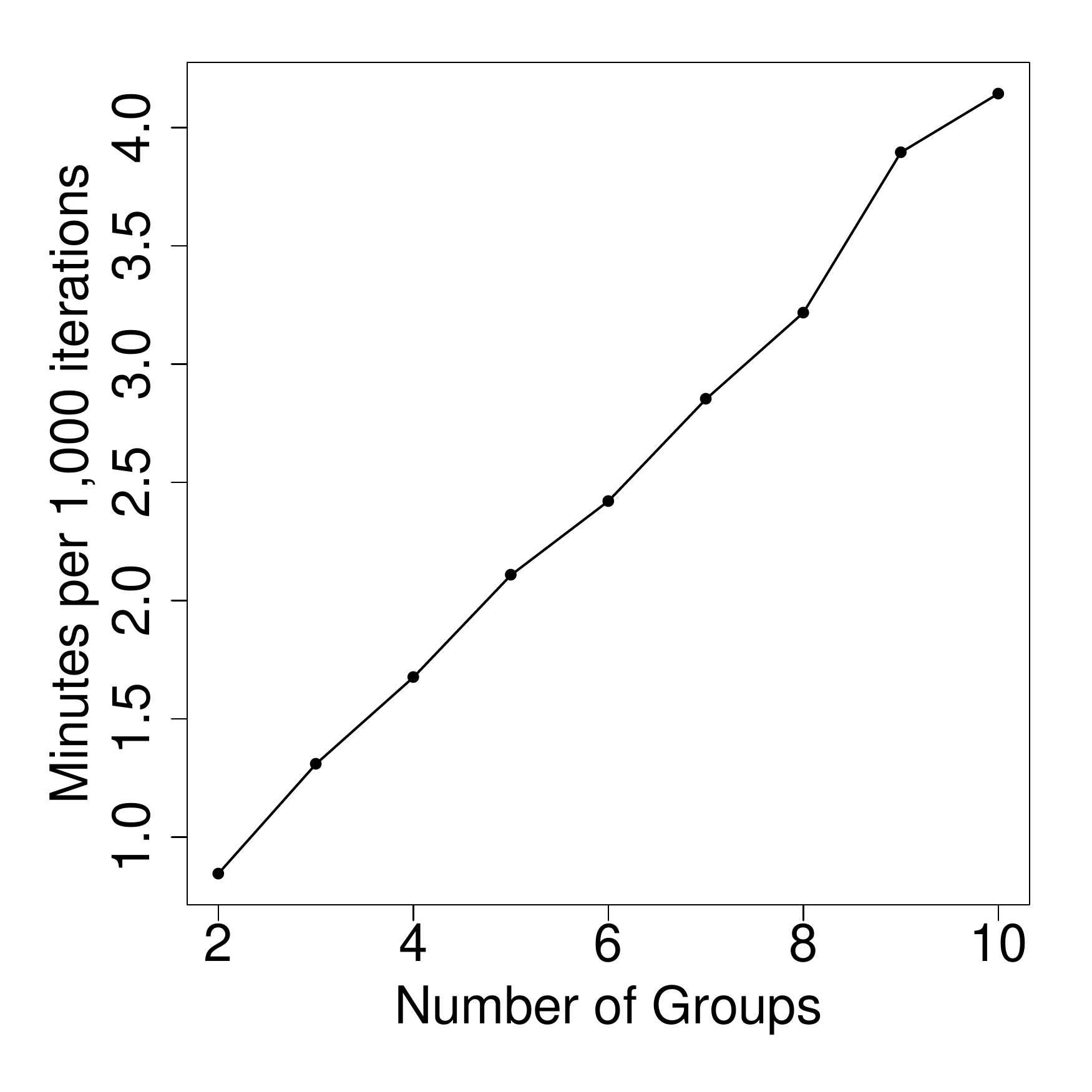}}
\caption{CPU time for 1,000 iterations of sampling. The left plot shows the CPU time for single graph with increasing node size, we compared our method (B-NS) with Bayesian Graphical Lasso (B-GLASSO) \citep{wang2012bayesian}. The right plot shows the CPU time with increasing number of graphs, where the node size was fixed to 100.}
\label{cpu_time}
\end{figure}

We evaluated the computational speed of the Bayesian variable selection procedure in the estimation of single GGM and multiple GGMs. For single GGM, we compared our method (B-NS) with Bayesian Graphical Lasso (B-GLASSO) \citep{wang2012bayesian} in Figure \ref{node}. Our algorithm took $0.5$ and $4.5$ minutes to generate 1,000 iterations for $p=100$ and $p=200$, and B-GLASSO took $1.6$ and $17.9$ minutes. We also evaluated the speed of our algorithm for the joint estimation of multiple graphs, where $n$ and $p$ were both fixed to 100. The CPU time was roughly linear as the number of graphs increased (Figure \ref{nsam}). All computations presented in Figure \ref{cpu_time} were implemented on a dual-core CPU 2.4 GHz laptop running OS X 10.9.5 using MATLAB 2014a. The computational cost of our algorithm is $O(p^3)$. When $p=500$, for single GGM, our algorithm took $\sim 60$ minutes for 1,000 iterations. In all the previous examples, we did not implement parallel computing. The computational time may be substantially reduced when the method is implemented in parallel by multicore processors (data not shown). Even for larger $p$ ($p\geq$500), our method may still be applicable if parallel computing is implemented.

\section{Application to the human brain gene expression dataset}
Next we apply our method to the human brain gene expression microarray dataset \citep{kang2011spatio}. In the dataset, the expression levels of $17,568$ genes were measured in 16 brain regions across 15 time periods. The time periods are not evenly spaced over time and each time period represents a distinct stage of brain development. Because of the small sample size, we incorporated the MRF model as in equation (\ref{mrf}) and did not consider more complex extensions: the brain regions were treated as parallel and the time periods were treated as discrete variables from $1$ to $15$. We excluded the data from time periods $1$ and $2$ in our analysis because they represent very early stage of brain development, when most of the brain regions sampled in future time periods have not differentiated. We also excluded the data where the number of replicates is less than or equal to $2$ (since a perfect line can be fitted with two data points), this step further removes $8$ groups of data: (brain region) ``MD'' (in time period) $4$, ``S1C'' $5$, ``M1C'' $5$, ``STR'' $10$, ``S1C'' $11$, ``M1C'' $11$, ``STR'' $11$ and ``MD" $11$. The number of replicates varies across brain regions/time periods and the number is less than $10$ in general, with a few exceptions. We studied the network of $7$ genes. These $7$ genes are high confidence genes that have been implicated in their roles for Autism Spectrum Disorders (ASD): GRIN2B, DYRK1A, ANK2, TBR1, POGZ, CUL3, and SCN2A\citep{willsey2013coexpression}. ASD is a neurodevelopment disorder that affects the brain and have an early onset in childhood. With a good understanding on the networks of the $7$ ASD genes across brain regions and time periods, we hope to gain insight into how these genes interact spatially and temporally to yield clues on their roles in autism etiology. The posterior mean and standard deviation for $\eta_s$ were $0.56$ and $0.51$, respectively. The posterior mean and standard deviation for $\eta_t$ were $0.95$ and $0.63$, respectively. The estimated model parameters suggest moderate sharing of network structure across brain regions and time.

\begin{figure}[h!]
\centering
\includegraphics[width=1\textwidth]{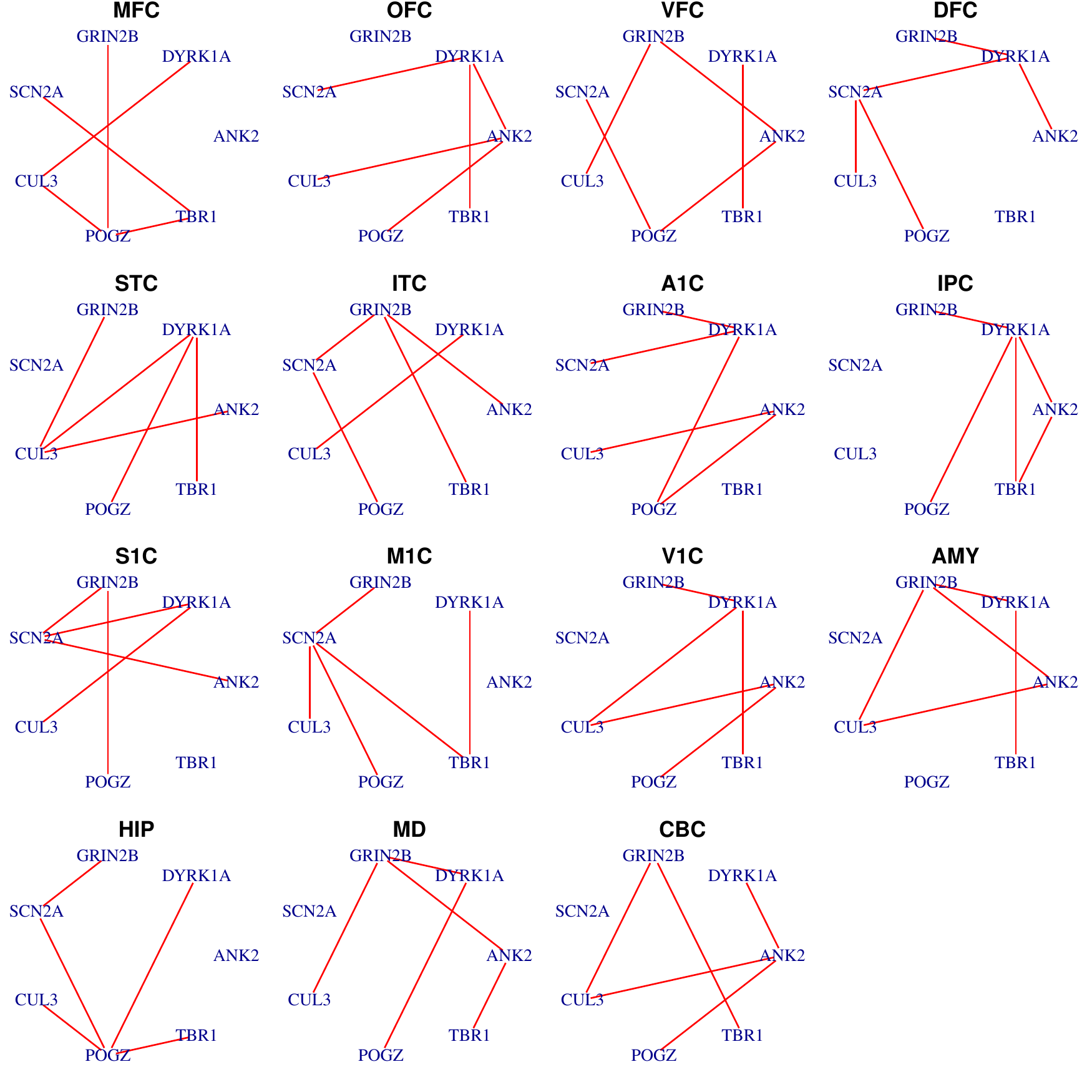}
\caption{The estimated graphs for all brain regions except ``STR'' in time period 10. Each graph corresponds to one brain region. Period 10 corresponds to early childhood ($1$ years $\leq$ age $\leq$ $6$ years), corresponding to the period of autism onset.}
\label{ASD_period10}
\end{figure}

\begin{figure}[h!]
\centering
\includegraphics[width=1\textwidth]{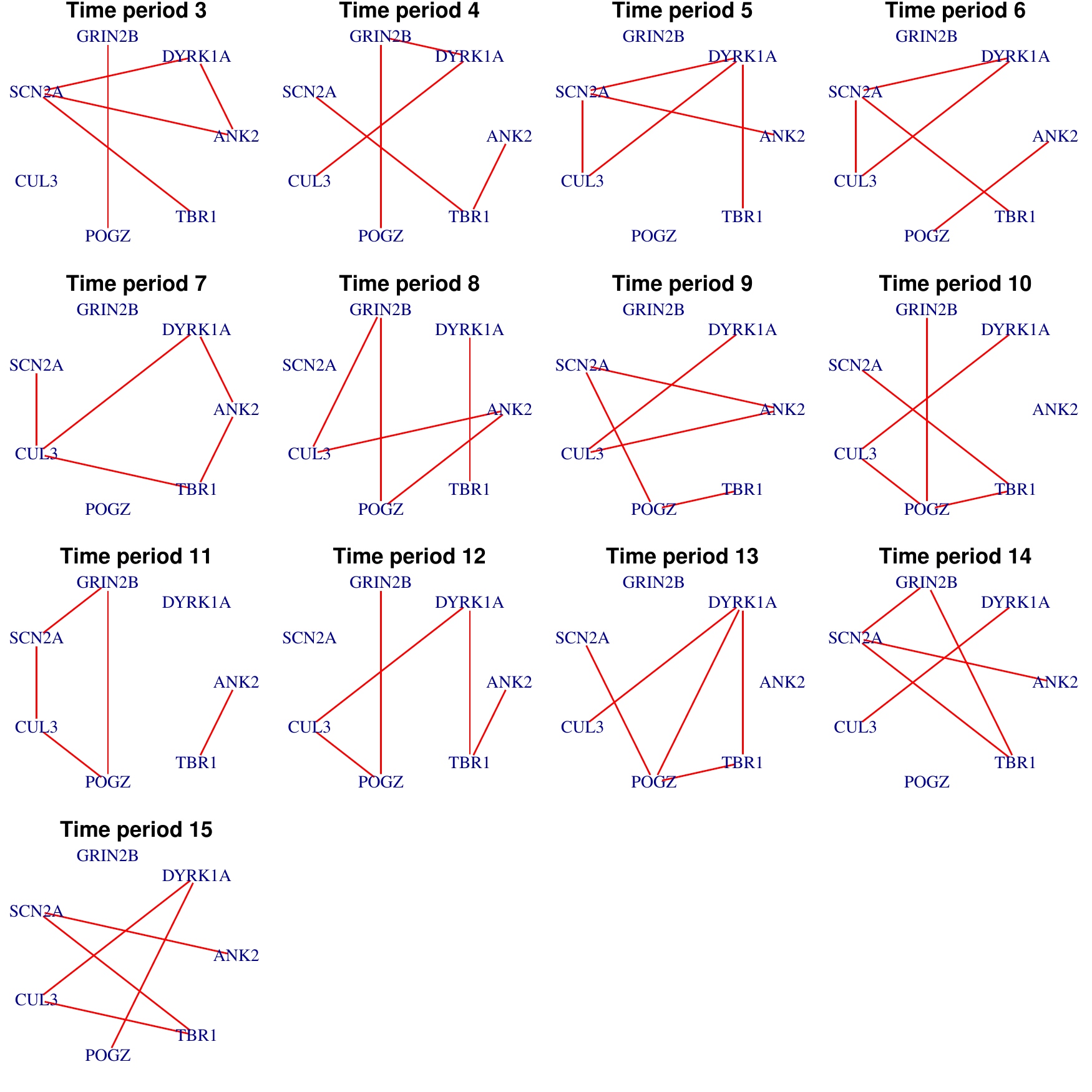}
\caption{The estimated graphs for brain region ``MFC'' across time. }
\label{ASD_MFC}
\end{figure}

For each graph, we selected the top $5$ edges with the highest marginal posterior probabilities. Time period 10 corresponds to early childhood ($1$ years $\leq$ age $\leq$ $6$ years), which is the typical period that patients show symptoms of autism. The graphs for all brain regions except ``STR'' (excluded data) are shown in Figure \ref{ASD_period10}. Of particular interest are the genes that are connected with TBR1, which is a transcription factor that may directly regulate the expression of numerous other genes. The edge between TBR1 and DYRK1A is mostly shared among the brain regions (7 regions). DYRK1A is a protein kinase that may play a significant role in the signaling pathway regulating cell proliferation and may be involved in brain development \citep{di2015chromatin}. It may be interesting to check whether TBR1 physically binds to DYRK1A during brain development. The graphs for region ``MFC'' across time are demonstrated in Figure \ref{ASD_MFC}. Because of the limit of space, we only show the temporal dynamics for one brain region. There are moderate sharing of edges over time. Although the edge between TBR1 and DYRK1A is not present in time period 10, it is present in time periods 5, 8, 12 and 13. Further biological experiments are required to validate whether the interaction between TBR1 and DYRK1A changes over time or it is caused by the lack of power to identify true edges due to the small sample size.

\section{Conclusion}
In this paper, we proposed a Bayesian neighborhood selection procedure to estimate Gaussian Graphical Models. Incorporating the Markov Random Field prior, our method was extended to jointly estimating multiple GGMs in data with complex structures. Compared with the non-Bayesian methods, there is no tuning parameter controlling the degree of structure sharing in our model. Instead, the parameters that represent similarity between graphs are learnt adaptively from the data. Simulation studies suggest that incorporating the complex data structure in the jointly modeling framework would benefit the estimation. We implemented our method by a fast and efficient algorithm that are several-fold faster than arguably the fastest algorithm for Bayesian inference of GGMs. Applying our method to the human brain gene expression data, we identified some interesting connections in the networks of autism genes during early childhood. We also demonstrated the graph selection consistency of our procedure for the estimation of single graph. The Matlab code is available at \url{https://github.com/linzx06/Spatial-and-Temporal-GGM}. 


\bibliographystyle{agsm} 
\bibliography{ref}

\end{document}


\def\spacingset#1{\renewcommand{\baselinestretch}%
{#1}\small\normalsize} \spacingset{1}

\if0\blind
{
  \title{\bf Supplementary Materials for ``On Joint Estimation of Gaussian Graphical Models for Spatial and Temporal Data''}
  \maketitle
} \fi

\if1\blind
{
  \bigskip
  \bigskip
  \bigskip
  \begin{center}
    {\LARGE\bf Title}
\end{center}
  \medskip
} \fi

\spacingset{1.45}
\section{The posterior sampling procedure}
For the estimation of multiple graphs, the steps for updating $\bm{\beta}$ and $\bm{\sigma}$ are the same as that for single graph, and thus we only show the update for single graph.

\subsection{Updating of $\bm{\beta}$}
The rows in $\bm{\beta}$ can be updated independently. Consider updating the $i$th row of $\bm{\beta}$. Let $D_i$ be a $(p-1)\times(p-1)$ dimensional diagonal matrix, and the $j$th diagonal entry takes the following value:
\[
    (D_i)_{jj}=
\begin{cases}
    \sigma_i^2/(\tau_{i1})^2,& \text{if } (\bm{\gamma}_{i\Gamma_i})_j=1,\\
    \sigma_i^2/(\tau_{i0})^2,& \text{if } (\bm{\gamma}_{i\Gamma_i})_j=0.
\end{cases}
\]
The full conditional distribution of $\bm{\beta}_{i\Gamma_i}$ follows multivariate Gaussian:
\begin{equation}
\bm{\beta}_{i\Gamma_i} \mid . \sim \mathcal{N}((\mathbf{X}_{\Gamma_i}'\mathbf{X}_{\Gamma_i}+D_i)^{-1} \mathbf{X}_{\Gamma_i}'\mathbf{X}_{i}, \sigma_i^2 (\mathbf{X}_{\Gamma_i}'\mathbf{X}_{\Gamma_i}+D_i)^{-1}).
\end{equation}
\subsection{Updating of $\bm{\sigma}$}
The elements in $\bm{\sigma}=(\sigma_1,...,\sigma_p)$ can be updated independently. The full conditional distribution of $\sigma_i$ follows inverse Gamma distribution:
\begin{equation*}
\sigma_i \mid .
\sim IG((\nu_i+n)/2, (\lambda \nu_i + (\mathbf{X}_{i} - \mathbf{X}_{\Gamma_i}\bm{\beta}_{i\Gamma_i}' )'(\mathbf{X}_{i} - \mathbf{X}_{\Gamma_i}\bm{\beta}_{i\Gamma_i}' ) )/2).
\end{equation*}
As $\nu_i=0$, 
\begin{equation*}
\sigma_i \mid .
\sim IG(n/2, (\mathbf{X}_{i} - \mathbf{X}_{\Gamma_i} \bm{\beta}_{i\Gamma_i}' )' (\mathbf{X}_{i} - \mathbf{X}_{\Gamma_i} \bm{\beta}_{i\Gamma_i}') /2 ).
\end{equation*}

\subsection{Updating of $\bm{\gamma}$}
For single graph, the full conditional probability of $\gamma_{ij}$ can be shown to be:
\begin{equation*}
p(\gamma_{ij} \mid .) \propto p(\beta_{ij} \mid \gamma_{ij}) p(\gamma_{ij}), \text{ for } j \neq i.
\end{equation*}
If there is symmetric constraint, the upper diagonal elements of $\bm{\gamma}$ can be updated as:
\begin{equation*}
p(\gamma_{ij} \mid .) \propto p(\beta_{ij} \mid \gamma_{ij}) p(\beta_{ji} \mid \gamma_{ji}=\gamma_{ij}) p(\gamma_{ij}), \text{ for } j > i.
\end{equation*}
Then let $\gamma_{ji}=\gamma_{ij}$.

For multiple graphs, the full conditional probability of $\gamma_{btij}$ can be shown to be:
\begin{equation*}
p(\gamma_{btij} \mid .) \propto p(\beta_{btij} \mid \gamma_{btij}) p(\gamma_{btij} \mid \bm{\gamma}_{\cdot \cdot ij}/\gamma_{btij}, \bm{\Phi} ), \text{ for } j \neq i.
\end{equation*}
If there is symmetric constraint, the full conditional distribution will be:
\begin{equation*}
p(\gamma_{btij} \mid.) \propto p(\beta_{btij} \mid \gamma_{btij}) p(\beta_{btji} \mid \gamma_{btji}=\gamma_{btij}) p(\gamma_{btij} \mid \bm{\gamma}_{\cdot \cdot ij}/\gamma_{btij}, \bm{\Phi} ), \text{ for } j > i.
\end{equation*}
Then let $\gamma_{btji}=\gamma_{btij}$.

\subsection{Updating of $\bm{\Phi}$}
$\bm{\Phi}=\{ \eta_1, \eta_s, \eta_t\}$. $\eta_1$ is prefixed. Then we have:
\begin{equation*}
p(\eta_s, \eta_t \mid .) \propto p(\eta_s, \eta_t)p(\bm{\gamma} \mid \bm{\Phi})=p(\eta_s)p(\eta_t)\prod_i\prod_{j \in \Gamma_i}p(\bm{\gamma}_{\cdot \cdot ij}\mid \bm{\Phi}),
\end{equation*}
where $p(\eta_s)$ and $p(\eta_t)$ represent the uniform priors on $\eta_s$ and $\eta_t$. The normalizing constant in $p(\bm{\gamma}_{\cdot \cdot ij}\mid \bm{\Phi})$ is generally not tractable as one has to sum over all $2^{|B|+|T|}$ possible configurations of $\bm{\gamma}_{\cdot \cdot ij}$. We approximate $p(\bm{\gamma}_{\cdot \cdot ij}\mid \bm{\Phi})$ by pseudolikelihood \citep{besag1986statistical}:
\begin{equation*}
p(\bm{\gamma}_{\cdot \cdot ij}\mid \bm{\Phi}) \approx \prod_{b \in B} \prod_{t \in T} p(\gamma_{btij} \mid \bm{\gamma}_{\cdot \cdot ij}/\gamma_{btij}, \bm{\Phi} ).
\end{equation*}
The Metropolis-Hastings (MH) algorithm was implemented to update $\eta_s$ and $\eta_t$.

\subsection{Computation}
When $p$ is large, the most computationally intensive step is updating $\bm{\beta}$, which involves the inversion of $(p-1) \times (p-1)$ matrices. The explicit matrix inversion can be avoided by first performing Cholesky decomposition. Consider updating $\bm{\beta}_{i\Gamma_i}$ by sampling from a multivariate Gaussian distribution with mean $(\mathbf{X}_{\Gamma_i}'\mathbf{X}_{\Gamma_i}+D_i)^{-1} \mathbf{X}_{\Gamma_i}'\mathbf{X}_{i}$ and covariance matrix $\sigma_i^2 (\mathbf{X}_{\Gamma_i}'\mathbf{X}_{\Gamma_i}+D_i)^{-1}$. By Cholesky decomposition, $\mathbf{X}_{\Gamma_i}'\mathbf{X}_{\Gamma_i}+D_i=R'R$, where $R$ is a upper-triangular matrix. The computational cost for Cholesky decomposition is $(p-1)^3/3$ and it is faster and more stable than direct matrix inversion. Next we sample $Z$ from $\mathcal{N}(0, I)$ and $(\mathbf{X}_{\Gamma_i}'\mathbf{X}_{\Gamma_i}+D_i)^{-1} \mathbf{X}_{\Gamma_i}'\mathbf{X}_{i} + \sigma_i R^{-1}Z$ follows the desired distribution of $\bm{\beta}_{i\Gamma_i}$.
\begin{equation*}
(\mathbf{X}_{\Gamma_i}'\mathbf{X}_{\Gamma_i}+D_i)^{-1} \mathbf{X}_{\Gamma_i}'\mathbf{X}_{i} + \sigma_i R^{-1}Z=R^{-1}( (R^{-1})'\mathbf{X}_{\Gamma_i}'\mathbf{X}_{i}+ \sigma_i Z).
\end{equation*}
$(R^{-1})'\mathbf{X}_{\Gamma_i}'\mathbf{X}_{i}$ and subsequently $R^{-1}( (R^{-1})'\mathbf{X}_{\Gamma_i}'\mathbf{X}_{i}+  \sigma_i Z)$ can be solved by forward and backward substitution and the computational cost is $O((p-1)^2)$.\\
Note that our model enables parallel computing in two levels: 1. for the estimation of a single graph, the rows in $\bm{\beta}$ can be updated in parallel independently; 2. for the joint estimation of multiple graphs, the matrix $\bm{\beta}$ for each graph can be updated in parallel independently. When multicore processors are available, parallel computing will result in substantial gain in computational speed.

\section{Choosing the hyperparameters}
\subsection{Choosing $\tau_1$ and $\delta$}
\begin{figure} 
\centering
\subfloat[$n=50$, sparsity$\sim 0.05$]{
\includegraphics[width=0.34\textwidth]{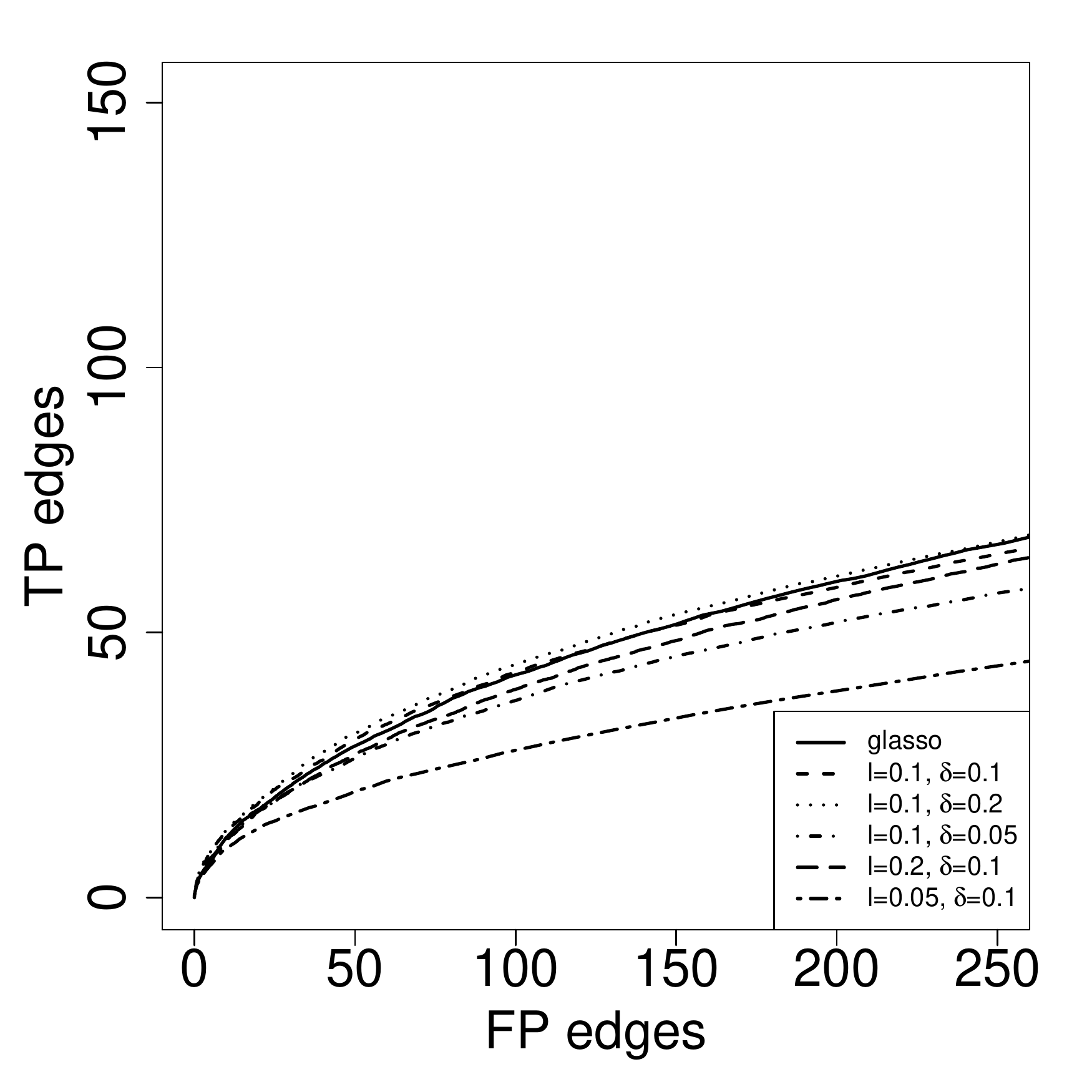}}
\subfloat[$n=150$, sparsity$\sim 0.05$]{
\includegraphics[width=0.34\textwidth]{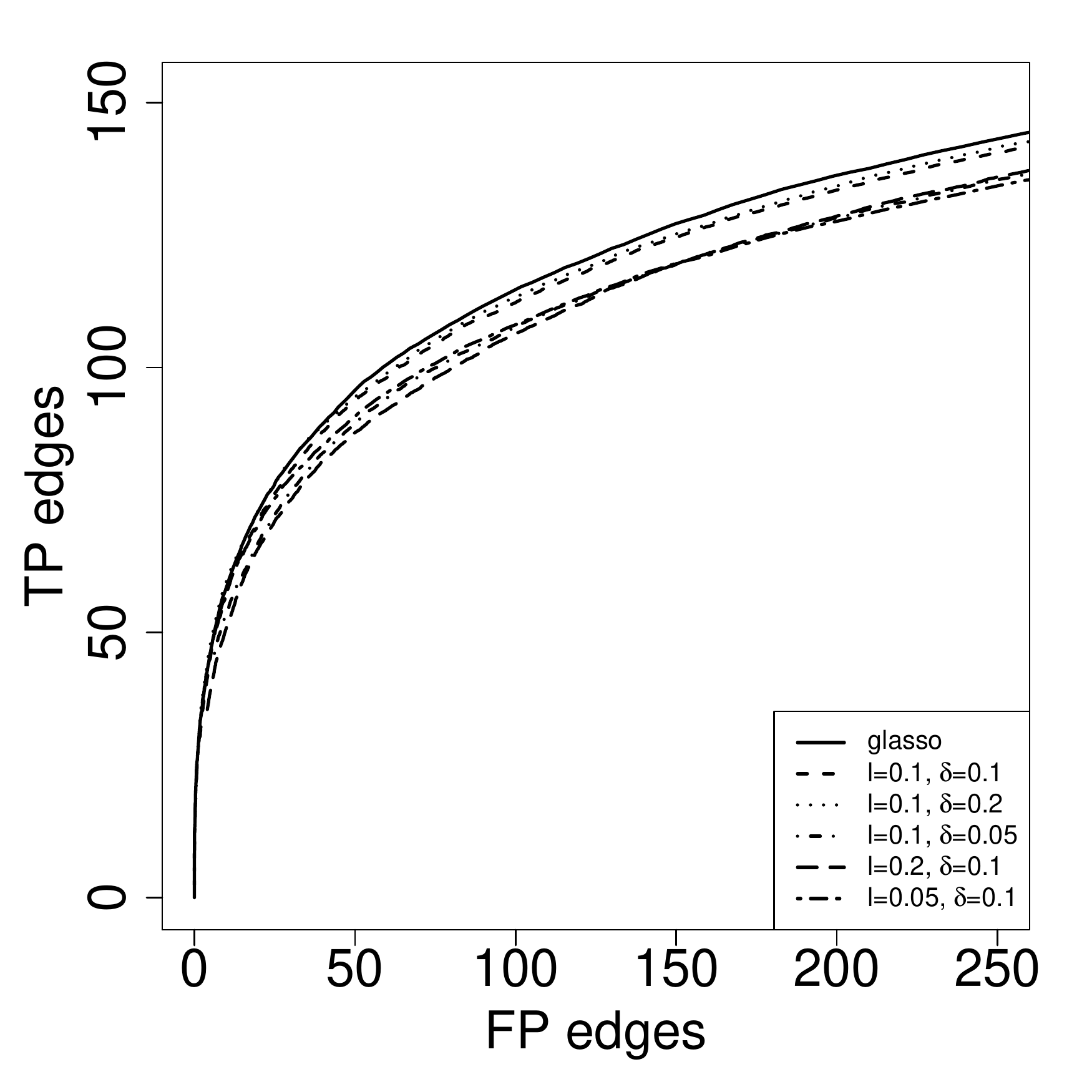}} \\
\subfloat[$n=50$, sparsity$\sim 0.1$]{
\includegraphics[width=0.34\textwidth]{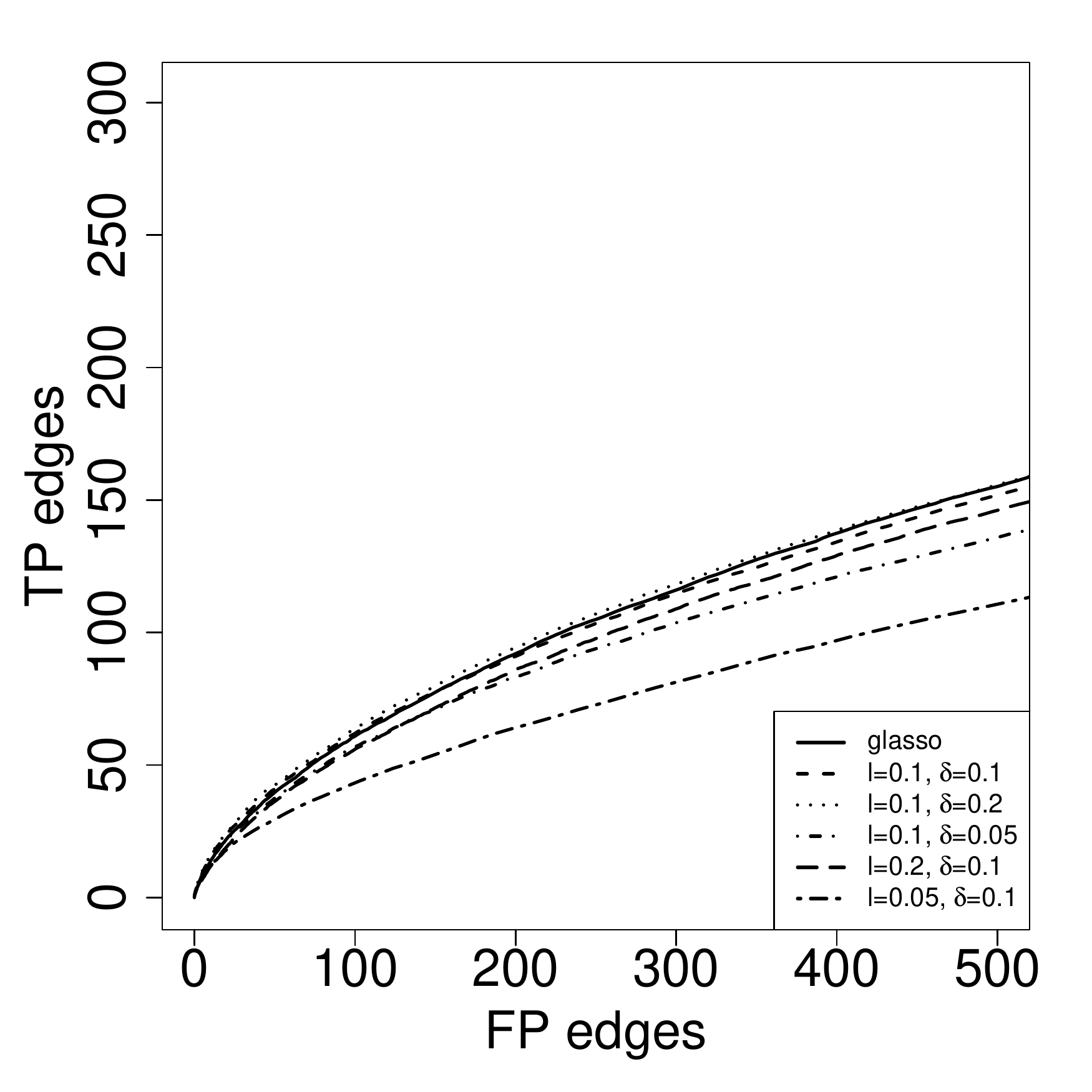}}
\subfloat[$n=150$, sparsity$\sim 0.1$]{
\includegraphics[width=0.34\textwidth]{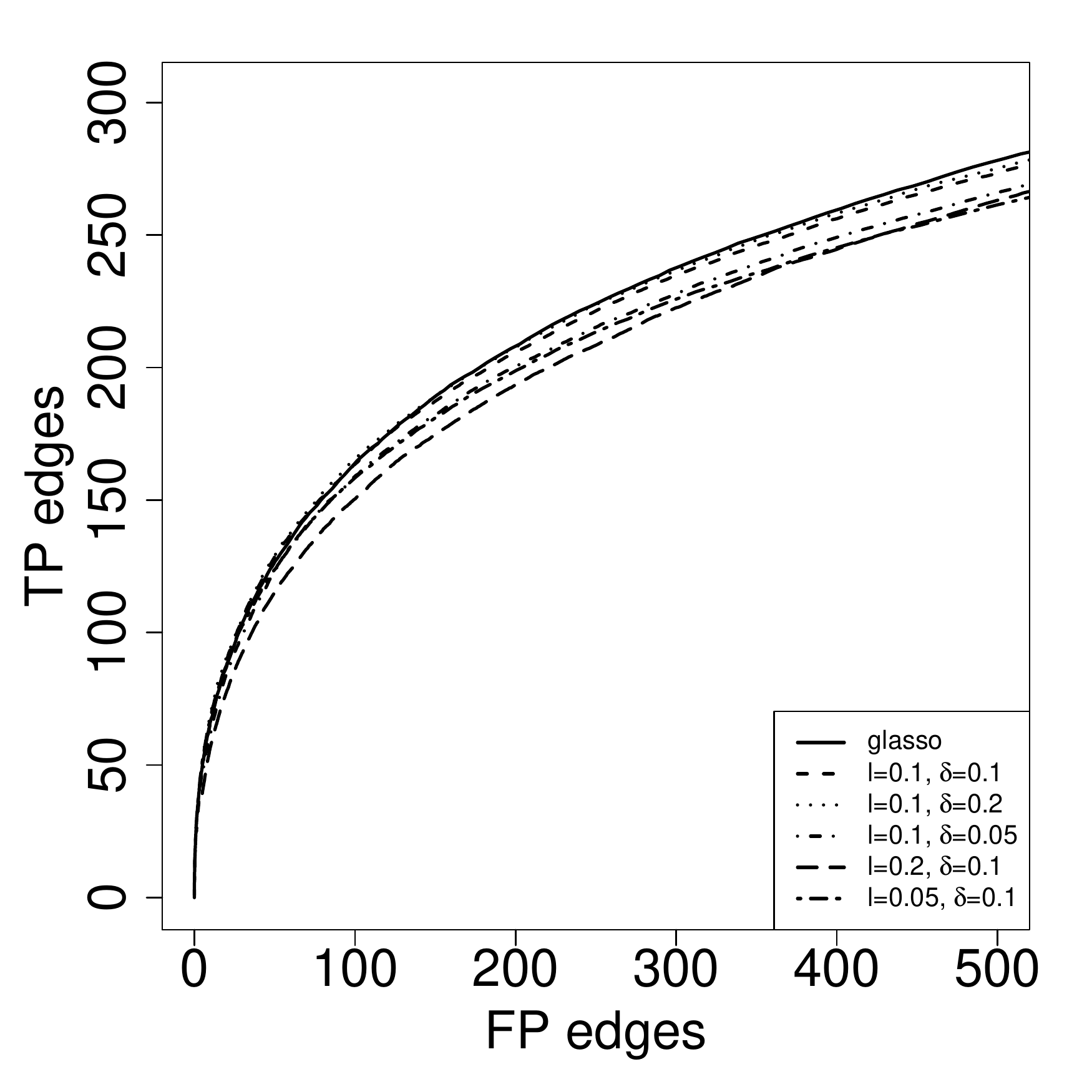}}
\caption{Performance of the model for different $l$ and $\delta$ (single graph). The curves represent the average of 100 independent runs.}
\label{change_rc}
\end{figure}

We performed diagnosis for $\tau_1$ and $\delta$ in single graph simulations. We simulated random graphs with sparsity$\sim 0.05$ or $0.1$, $p=100$, and $n=50$ or $150$. 
Let $s_i$ be the standard deviation of $X_i$ and let $\tau_{i1}=l s_i$. We fixed $q$ to the corresponding sparsity level and varied $l$ and $\delta$ (Figure \ref{change_rc}). When $n=150$, the performance did not vary much when $l$ and $\delta$ varied. We set $l=0.1$ and $\delta=0.1$ in the simulation of multiple graphs. When $n<p$, numerical issues can arise as the entries in $D_i$ will be close to $0$ and $\mathbf{X}_{\Gamma_i}'\mathbf{X}_{\Gamma_i}+D_i$ will become singular. The numerical issue can be avoided by decreasing $q$ or $l$. For the real data example, we set $l=0.1n/p$ if $n<p$.

\subsection{Choosing the Bernoulli parameter $q$}
\begin{figure} 
\centering
\subfloat[$n=50$, sparsity$\sim 0.05$]{
\includegraphics[width=0.34\textwidth]{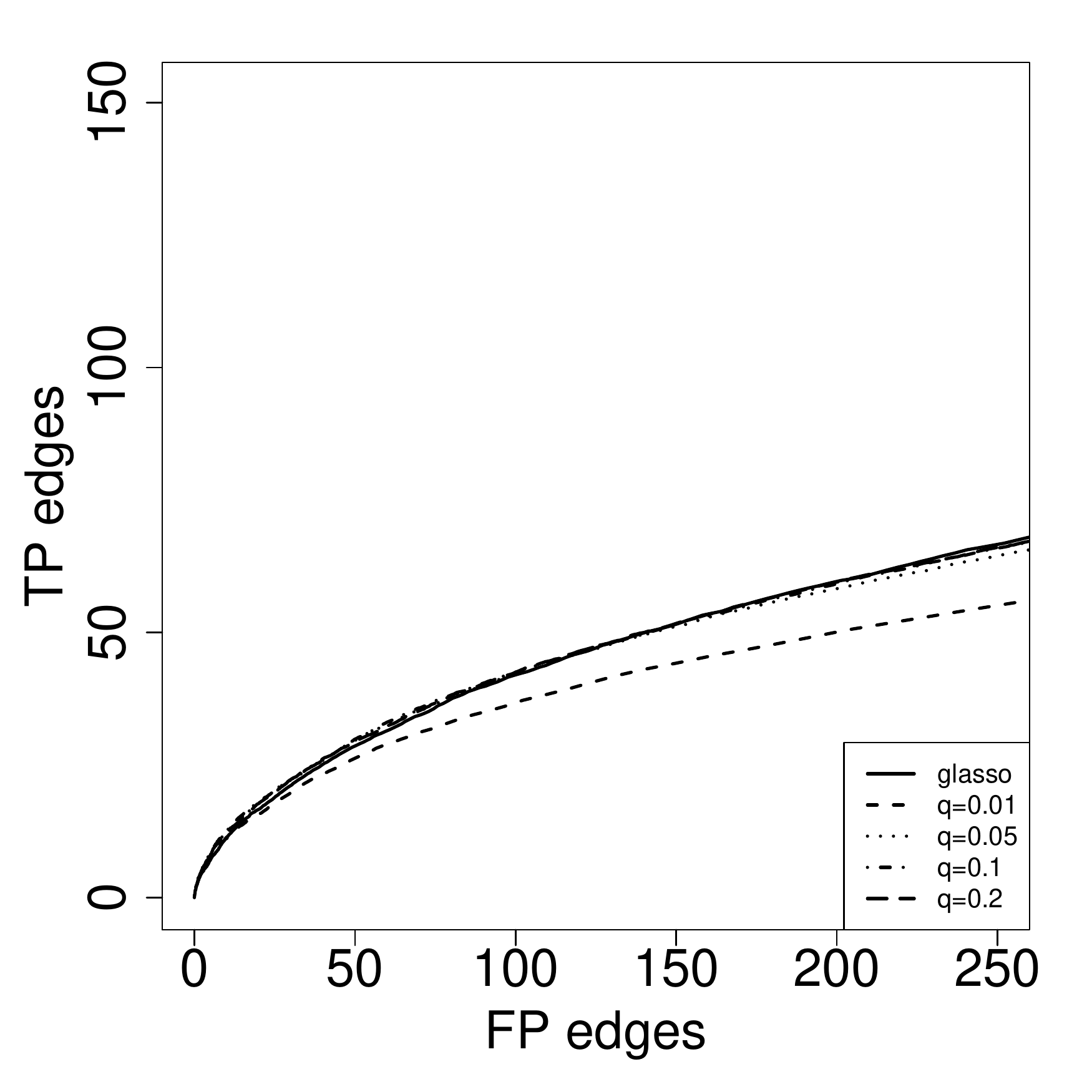}}
\subfloat[$n=150$, sparsity$\sim 0.05$]{
\includegraphics[width=0.34\textwidth]{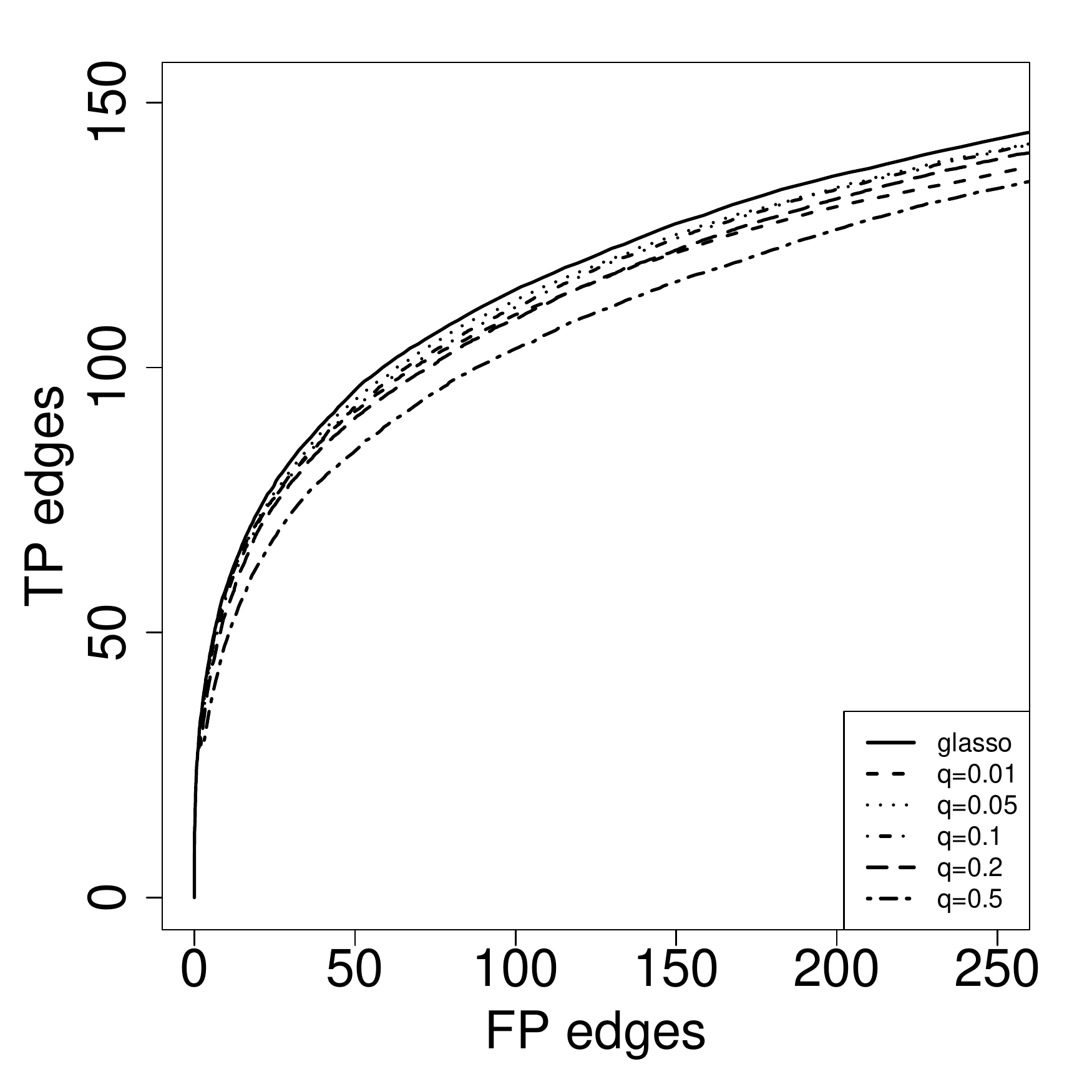}} \\
\subfloat[$n=50$, sparsity$\sim 0.1$]{
\includegraphics[width=0.34\textwidth]{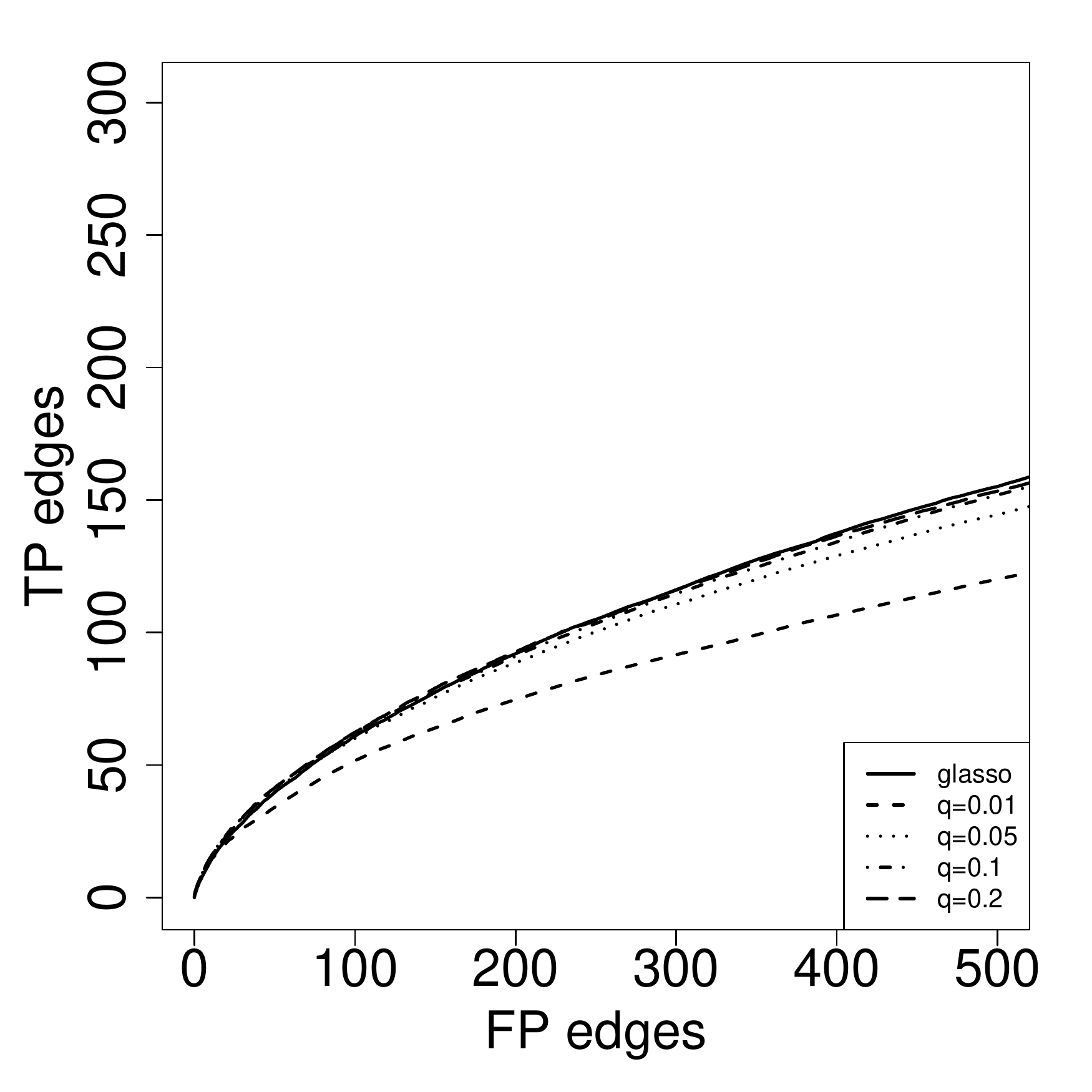}}
\subfloat[$n=150$, sparsity$\sim 0.1$]{
\includegraphics[width=0.34\textwidth]{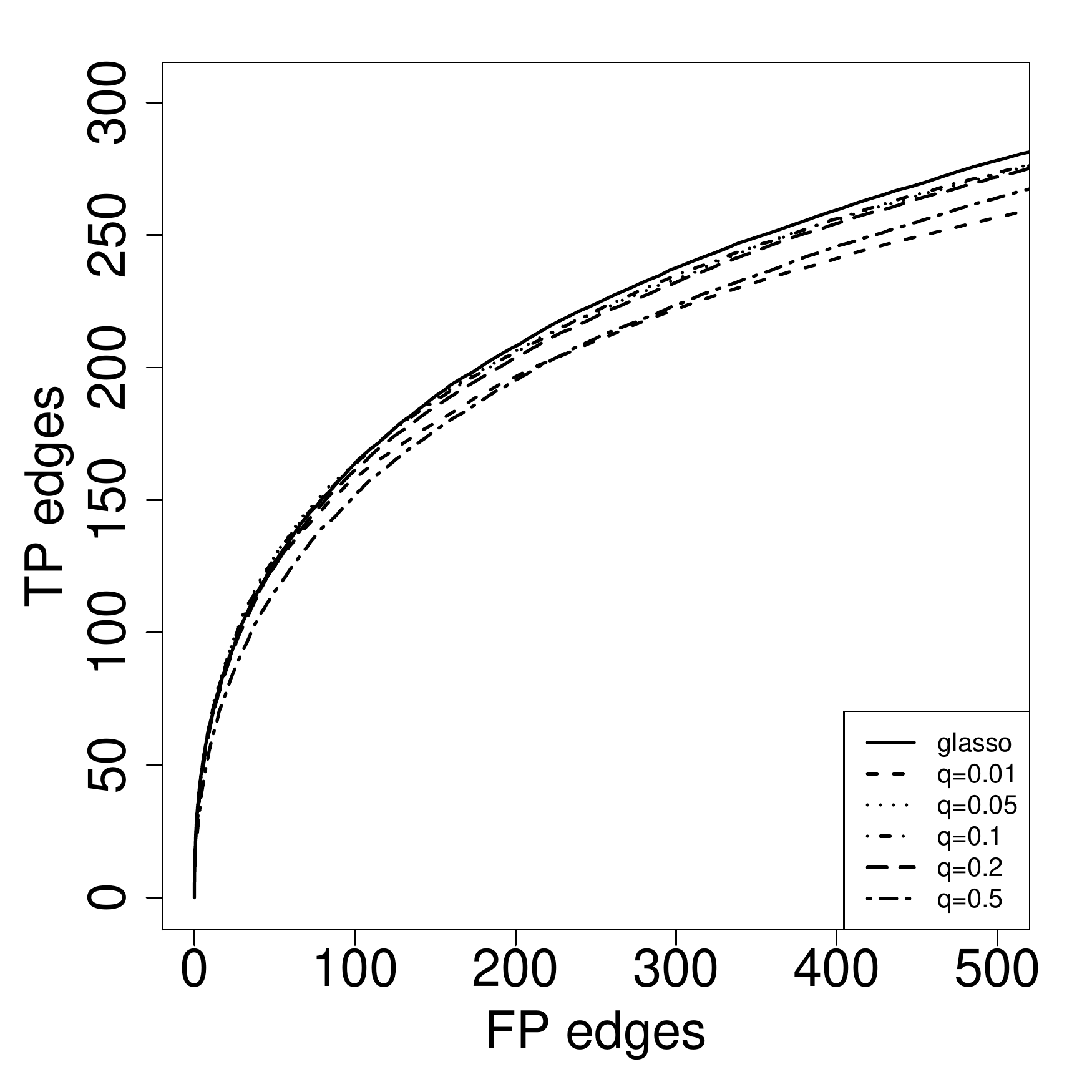}}
\caption{Performance of the model for different $q$ (single graph). The curves represent the average of 100 independent runs.}
\label{varyq}
\end{figure}

We simulated random graphs with sparsity$\sim 0.05$ or $0.1$, $p=100$, and $n=50$ or $150$. We fixed $l=0.1$, $\delta=0.1$ and varied $q$. The graph selection consistency requires $q=p^{-1}$. In our simulation examples, the posterior probabilities are sensitive to $q$, but the ROC is quite robust to the choice of $q$ (Figure \ref{varyq}). 

\subsection{Choosing $\eta_1$ in the MRF prior}
\begin{figure} 
\centering
\subfloat[Sparsity$\sim 0.05$]{
\includegraphics[width=0.34\textwidth]{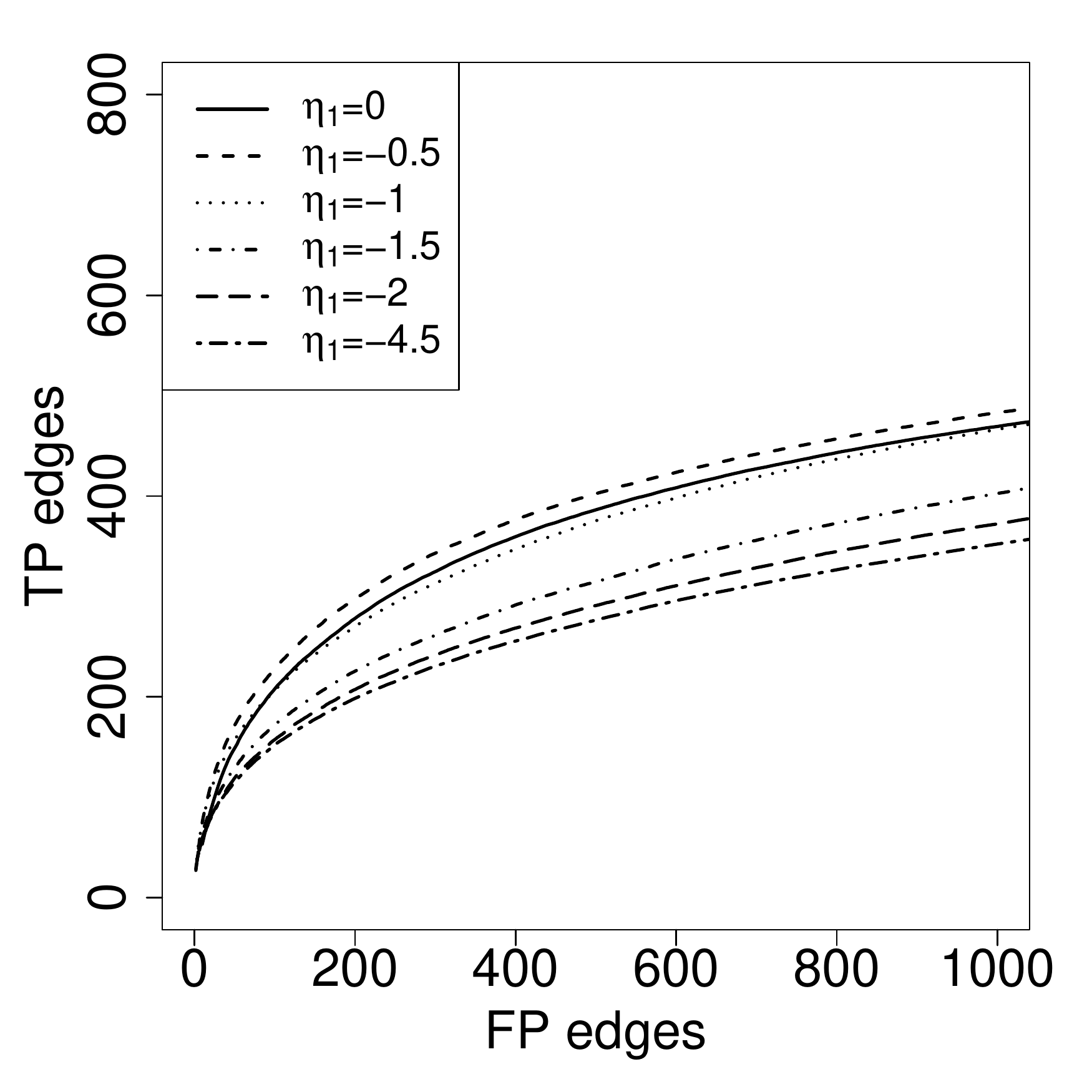}}
\subfloat[Sparsity$\sim 0.1$]{
\includegraphics[width=0.34\textwidth]{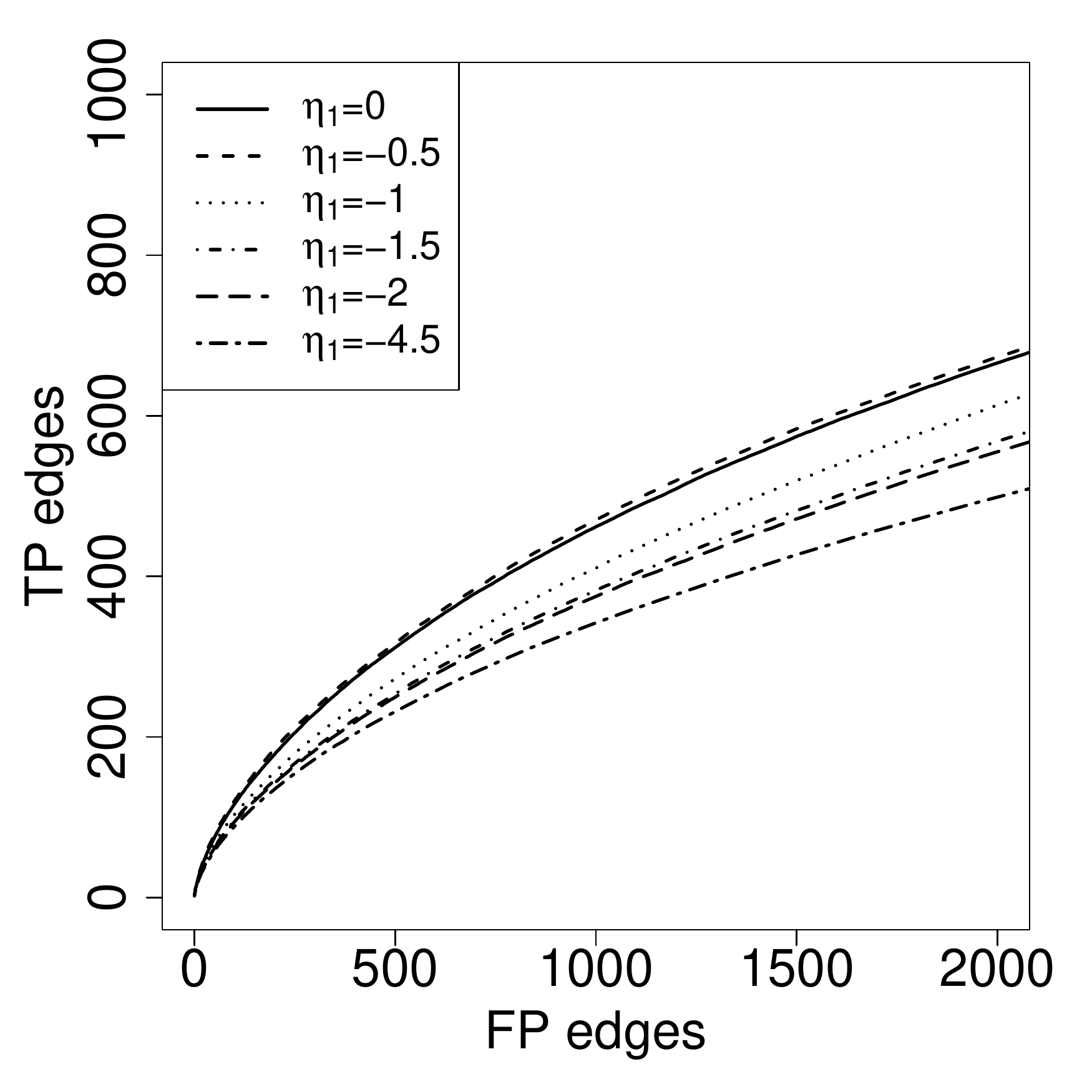}}
\caption{Performance of the model for different $\eta_1$. The curves represent the average of 100 independent runs.}
\label{varyeta1}
\end{figure}

In the simulation studies, the performance of our model was quite sensitive to the choice of $\eta_1$. This may be caused by the fact that we used pseudolikelihood for approximation. We simulated three random graphs same as that in Section 5.1 in the main text, where $p=100$, $n=150$ and change$=0.2$. We found that when $\eta_1=-0.5$, our proposed model has reasonable performance (Figure \ref{varyeta1}). Therefore we fix $\eta_1=-0.5$ for both the simulation studies and the real data example.

\section{Proofs}
Let $Y^i = \X_i$ and $Z^i = \X_{V\setminus\{i\}}$. Since the proof is independent of the node index, in the sequel we suppress the script $i$. We also suppress $v$ and $K$ from the notation of $m(v), \lambda_{m}(v)$, and $\Delta(K)$. Throughout, we use $c', s'$ and $w'$ to denote generic positive constants that can take different values each time they appear.

Define the Bayes factor of model $k$ with respect to the true model $t$ as
$$BF(k, t) = \frac{P(\bfgamma = k \mid Y, Z, \sigma^2)}{P(\bfgamma = t \mid Y, Z, \sigma^2)}.$$
Let $D_k = {\rm diag}\{\tau_{1n}^{-2} k + \tau_{0n}^{-2} (1 - k)\}$ and $\tilde{R}_k = Y'\{I - Z(D_k + Z'Z)^{-1}Z'\} Y$.
\begin{lemm}\label{lemma1}
Under the conditions of Theorem 1, for any $k \neq t$, we have
$$BF(k, t) \leq w' \{n \tau_{1n}^2 \lambda_m (1 - \phi_n)\}^{-(r_k^* - r_t) / 2} \lambda_m^{-|t \wedge k^c| / 2} q_n^{|k| - |t|} \exp\{-(\tilde{R}_k - \tilde{R}_t) / (2\sigma^2)\},$$
where $r_k = rank(Z_k), r_k^* = r_k \wedge m$, and $\phi_n = o(1)$ uniformly in $k$.
\end{lemm}
{\it Proof of Lemma \ref{lemma1}}. This is Lemma 4.1 of \citet{narisetty2014bayesian}, and we omit the proof.

Let $R_k = Y' \{I - Z_k(\tau_{1n}^{-2}I + Z_k' Z_k)^{-1}Z_k'\}Y$ and $R_k^* = Y'(I - P_k)Y$.

\begin{lemm}\label{lemma2}
Assume the conditions of Theorem 1. Let $j$ and $k$ be models such that $j \subseteq k$. We have
$$(R_j - R_k) (1 - w_n)(1 - \xi_n)^2 \leq \tilde{R}_j - \tilde{R}_k \leq (R_j - R_k)(1 - \xi_n)^{-1},$$
where $w_n = o(1)$ uniformly in $k$, and $\xi_n = n \tau_{0n}^2 \lambda_{\max}(Z'Z/n) = o(1)$.
\end{lemm}
{\it Proof of Lemma \ref{lemma2}}. This is Lemma A.1 (iii) of \citet{narisetty2014bayesian}, and we omit the proof.


\begin{lemm}\label{lemma3}
Assume the conditions of Theorem 1. For any $\varepsilon > 0$ and any sequence $g_n$ such that $\log p_n \preceq g_n$, we have
$$P(|R_t^* - n\sigma^2| > \varepsilon n \sigma^2) \leq \exp(-c' n)$$
and
$$P(R_t - R_t^* > g_n) \leq \exp(- c' n \tau_{1n}^2 g_n).$$
\end{lemm}
{\it Proof of Lemma \ref{lemma3}}. This is Lemma A.2 of \citet{narisetty2014bayesian}, and we omit the proof.


{\it Proof of Theorem 1}. Following \citet{narisetty2014bayesian}, we divide the set of possible incorrect models into four subsets
$$M_1 = \{k: r_k > m\},$$
$$M_2 = \{k: k \supseteq t, k \neq t, r_k \leq m\},$$
$$M_3 = \{k: k \not\supseteq t, K |t| < r_k \leq m\},$$
and
$$M_4 = \{k: k \not\supseteq t, r_k \leq K|t|\}.$$
We shall prove $\sum_{k \in M_u} BF(k,t) \stackrel{P}{\longrightarrow} 0$ for each $u=1,2,3$, and $4$.

{\bf Unrealistically large models.} We first consider the models in $M_1$, which correspond to all the models containing at least $m$ linearly independent covariates. Note that $M_1$ is empty if $p_n \leq n/\log(p^{2+\nu}_n)$. Thus, we assume for the moment that $p_n >  n/\log(p^{2+\nu}_n)$. Then we have $r^*_k = m = n/\log(p^{2+\nu}_n)$ for all $k \in M_1$. For any $s>0$,
\begin{eqnarray*}
& & P[\cup_{k\in M_1} \{\tilde{R}_t - \tilde{R}_k > (1+2s)\sigma^2 n\}]\\ &\leq& P\{\tilde{R}_t > (1+2s)\sigma^2n\}\\
&\leq&  P\{R_t > (1+2s)\sigma^2 n\} \\
&\leq& P\{R^*_t > (1+s) \sigma^2 n\} + P(R_t - R^*_t > s\sigma^2 n).
\end{eqnarray*}
By Lemma \ref{lemma3}, 
\begin{eqnarray*}
P[\cup_{k\in M_1} \{\tilde{R}_t - \tilde{R}_k > (1+2s)\sigma^2 n\}] \leq 2 \exp(-c'n).
\end{eqnarray*}

Restricting to the event $\{\tilde{R}_t - \tilde{R}_k \leq (1+2s)\sigma^2 n, \forall k \in M_1\}$, Lemma \ref{lemma1} and Condition (G) give
\begin{eqnarray*}
\sum_{k \in M_1} BF(k, t) &\preceq& \{n \tau_{1n}^2 \lambda_m (1 - \phi_n)\}^{-(m - r_t) / 2}  q_n^{|k|-|t|} \lambda_m^{-|t|/2} \exp\{(1+2s)n/2\}\\ &\preceq& \sum_{k \in M_1} p^{-(1+\delta_2)(m - |t|)}_n q_n^{|k|-|t|} \lambda_m^{-|t|/2} \exp\{(1+2s)n/2\}.
\end{eqnarray*}
Note that $m = n/\log(p^{2+\nu}_n), |t| = o(m)$, and $v < \delta_2$. Then
\begin{eqnarray*}
&& \sum_{k \in M_1} BF(k, t)\\ &\preceq& \sum_{k \in M_1} \exp\{-n(1+\delta_2)/(2+\delta_2)\} q^{|k|-|t|}_n \lambda_m^{-|t|/2} \exp\{(1+2s)n/2\}\\
&=& \sum_{k \in M_1} \exp\{-n(1+\delta_2)/(2+\delta_2) + (1+2s)n/2\} q^{|k|-|t|}_n \lambda_m^{-|t|/2}.
\end{eqnarray*}
By Conditions (B) and (G), we have
\begin{eqnarray*}
&&\sum_{k \in M_1} BF(k, t)\\ &\preceq& \exp\{-n(1+\delta_2)/(2+\delta_2) + (1+2s)n/2\}p^{s' |t|}_n \sum_{k\in M_1}q^{|k|}_n \\
&\preceq& \exp\{-n(1+\delta_2)/(2+\delta_2) + (1+2s)n/2\}p^{s' |t|}_n \sum^{p_n}_{|k|=m + 1} \binom{p_n}{|k|} q^{|k|}_n\\
&\preceq& \exp\{-n(1+\delta_2)/(2+\delta_2) + (1+2s)n/2\}p^{s' |t|}_n (1+q_n)^{p_n}.
\end{eqnarray*}
By Conditions (A) and (B), $p_n \log(1 + q_n) \sim p_n^{\alpha} = o(n)$, and by Condition (D), $|t| \log(p_n) = o(n)$. Hence
\begin{eqnarray*}
\sum_{k \in M_1} BF(k, t) \preceq  \exp(-w'n)
\end{eqnarray*}
for some $w' > 0$, if $s$ satisfies $2s < \delta_2/(2+\delta_2)$. Therefore, with probability at least $1 - 2\exp(-c'n)$,
\begin{equation}\label{eq2}
\sum_{k \in M_1} BF(k, t) \rightarrow 0.
\end{equation}

{\bf Over-fitted models.} For $k \in M_2$, we have
\begin{eqnarray*}
R^*_t - R^*_k = Y' (P_k - P_t) Y= \|(P_k - P_t)Y\|^2_2 = \|(P_k - P_t)\epsilon\|_2^2 = \epsilon' (P_k- P_t) \epsilon,
\end{eqnarray*}
where $\epsilon \sim N(0, \sigma^2 I)$. Since $\epsilon' (P_k- P_t) \epsilon / \sigma^2$ follows the chi-squared distribution with $r_k - r_t$ degrees of freedom, for any $s > 0$ and $a > 1$, we have
\begin{eqnarray*}
&&P\{R^*_t - R^*_k > \sigma^2(2+ 2 sa) (r_k-r_t) \log p_n\}\\ &=& P\left\{\frac{\epsilon' (P_k- P_t) \epsilon}{\sigma^2}> (2+ 2 sa) (r_k-r_t) \log p_n\right\}\\ &=& P\left\{\frac{\chi^2_{r_k - r_t}}{r_k -r_t} > 1 + (2+2sa) \log p_n - 1\right\}\\
&\leq& \exp\{-(r_k-r_t)(1+ w sa)\log p_n\}
\end{eqnarray*}
for some $1/ a < w < 1$. Hence
\begin{eqnarray}\label{eq3}
& & P\{R^*_t - R^*_k > \sigma^2(2+ 2 sa) (r_k-r_t) \log p_n\} \leq p^{-(1+s)(r_k-r_t)}_n.
\end{eqnarray}

Consider $s > 1$ and $a > 1$. Define the event
\begin{eqnarray*}
A(k) = \{\tilde{R}_t - \tilde{R}_k > 2\sigma^2 (1+sa)(r_k -r_t) \log p_n\}.
\end{eqnarray*}
By Lemma \ref{lemma2}, we have, for some $1 < a' < a$,
\begin{eqnarray*}
A(k) &\subseteq& \{R_t-R_k > 2\sigma^2 (1+sa)(r_k -r_t)(1-\xi_n) \log p_n\} \\
&\subseteq& \{R_t-R_k > 2\sigma^2 (1+sa')(r_k -r_t)\log p_n\} \\
&\subseteq& \{{R_t-R_k^*} > 2\sigma^2 (1+sa')(r_k -r_t)\log p_n\}.
\end{eqnarray*}
For a fixed dimension $d > r_t$, consider the event $U(d) = \cup_{\{k \in M_2: r_k=d\}} A(k)$. Then, for any $1 < w' < a'$,
\begin{eqnarray}\label{eq4}
&&P\{U(d)\} \nonumber
 \\&\leq& P[ \cup_{\{k \in M_2: r_k=d\}} \{{R_t - R^*_k} > 2\sigma^2 (1+sa')(d-r_t)\log p_n\}] \nonumber\\
&\leq& P[ \cup_{\{k \in M_2: r_k=d\}} \{{R^*_t - R^*_k} > \sigma^2(2+2sw')(d-r_t)\log p_n\} ] \\
& & + P[R_t - R^*_t > \sigma^2 (2sa' - 2sw')  (d-r_t)\log p_n].\nonumber
\end{eqnarray}
The event $\{R^*_t - R^*_k > \sigma^2(2+2sw')(r_k-r_t)\log p_n\}$ depends only on the projection matrix $P_{k} - P_t$. When $r_t < r_k = d \leq m$, the cardinality of such projections is at most $(p_n - r_t)^{d-r_t} \leq p_n^{d-r_t}$. This, together with (\ref{eq3}), (\ref{eq4}) and Lemma \ref{lemma3}, yields
\begin{eqnarray}\label{eq5}
P\{U(d)\} &\leq& p^{-(1+s)(d-r_t)}_n p^{d-r_t}_n + \exp(-c'n\tau_{1n}^2\log p_n)\\
&\leq& 2  p^{-s(d-r_t)}_n.\nonumber
\end{eqnarray}
Hence
\begin{eqnarray*}\label{eq6}
P\left\{\cup_{d>r_t}U(d)\right\} \leq \sum_{d>r_t}P\{U(d)\} \leq 2  \sum_{d>r_t} p^{-s(d-r_t)}_n \leq c'{p^{-s}_n}. \end{eqnarray*}

Restricting to the event $\cap_{d>r_t} \{U(d)\}^c$, by Lemma \ref{lemma1} and Condition (G), we have
\begin{eqnarray*}
\sum_{k\in M_2} BF(k,t) &\preceq& \sum_{k\in M_2} \{n\tau^2_{1n} \lambda_m (1-\phi_n)\}^{-(r_k-r_t)/2} q^{|k|-|t|}_n p^{(1+sa)(r_k-r_t)}_n\\
&\preceq& \sum_{k\in M_2} \left(\frac{p^{1+sa}_n q_n}{p^{1+\delta_2}_n \vee \sqrt{n}}\right)^{r_k-r_t} \\
&\preceq& \left(1 + \frac{p^{1+sa}_n q_n}{p^{1+\delta_2}_n \vee \sqrt{n}}\right)^{p_n} - 1.
\end{eqnarray*}
Since $\delta_2 > 1 + \alpha$, there always exist $s > 1$ and $a > 1$ such that
\begin{eqnarray*}
p_n \log\left(1 + \frac{p^{1+sa}_n q_n}{p^{1+\delta_2}_n \vee \sqrt{n}}\right) \sim p_n \frac{p^{1+sa}_n q_n}{p^{1+\delta_2}_n \vee \sqrt{n}}  = \frac{p^{1+sa + \alpha}_n }{p^{1+\delta_2}_n \vee \sqrt{n}} = o(1).
\end{eqnarray*}
Therefore, with probability at least $1 - c'p_n^{-s}$ for some $s > 1$, we have
\begin{equation}\label{eq7}
\sum_{k \in M_2} BF(k, t) {\rightarrow} 0.
\end{equation}

{\bf Large models.} Models in $M_3$ do not contain one or more active covariates with dimension at least $K|t| + 1$. Since $D_k \geq D_{k \vee t}$, we have
\begin{eqnarray*}
&&\tilde{R}_k - \tilde{R}_{k \vee t}\\ &=& Y'\{I - Z(D_k + Z'Z)^{-1}Z'\} Y - Y'\{I - Z(D_{k \vee t} + Z'Z)^{-1}Z'\} Y\\
&=& Y'\{Z(D_{k \vee t} + Z'Z)^{-1}Z' - Z(D_k + Z'Z)^{-1}Z'\} Y \geq 0.
\end{eqnarray*}

Similar to the proof in the previous part, we define the event
\begin{eqnarray*}
B(k) &=& \left\{\tilde{R}_t - \tilde{R}_k > 2\sigma^2 (1+4s)(r_k-r_t) \log p_n\right\} \\
&\subseteq& \left\{\tilde{R}_t - \tilde{R}_{k\vee t} > 2\sigma^2 (1+4s)(r_k-r_t) \log p_n\right\} \\
&\subseteq& \{R_t - R_{k\vee t} > 2\sigma^2 (1+2s)(r_k-r_t)\log p_n\},
\end{eqnarray*}
and consider the union of such events $V(d) = \cup_{\{k \in M_3: r_k=d\}} B(k)$. For $s > 0$, we have
\begin{eqnarray*}
P\{V(d)\} &\leq& P[\cup_{\{k \in M_3: r_k=d\}}\{R_t - R_{k\vee t} > 2\sigma^2 (1+2s)(d-r_t)\log p_n\}] \\
&\leq& P[\cup_{\{k \in M_3: r_k=d\}}\{R_t^* - R_{k\vee t} > \sigma^2 (2+3s)(d-r_t)\log p_n\}]\\ & & +\ P\{R_t - R_{t}^* > \sigma^2 s(d-r_t)\log p_n\}\\
&\leq& P[\cup_{\{k \in M_3: r_k=d\}}\{R_t^* - R_{k\vee t} > \sigma^2 (2+3s)(d-r_t)\log p_n\}]\\ & & +\ \exp(-c'|t|n\tau_{1n}^2\log p_n).
\end{eqnarray*}
Further,
\begin{eqnarray*}
& & P\{R_t^* - R_{k\vee t} > \sigma^2 (2+3s)(r_k-r_t)\log p_n\} \\
&\leq& P\{R_t^* - R_{k\vee t}^* > \sigma^2 (2+3s)(r_k-r_t)\log p_n\} \\
&=& P\left\{\frac{\chi^2_{|k \wedge t^c|}}{|k \wedge t^c|} > 1 + \frac{(2+3s)(r_k-r_t)\log p_n}{|k \wedge t^c|} - 1\right\}  \\
&=& P\left\{\frac{\chi^2_{|k \wedge t^c|}}{|k \wedge t^c|} > 1 + \frac{(2+3s) (K - 1)\log p_n}{K} - 1\right\} \\
&\leq& \exp\{-|k \wedge t^c| (1 + s)\log p_n\} \\
&\leq& \exp\{-(r_k - r_t) (1 + s)\log p_n\}.
\end{eqnarray*}
Similar to (\ref{eq5}),
\begin{eqnarray*}
P\{V(d)\} \leq  p_n^{-(1+s)(d - r_t) + d} + \exp(-c'|t|n\tau_{1n}^2\log p_n) \leq 2 p_n^{-(1+s)(d - r_t) + d}.
\end{eqnarray*}
By Condition (E), there exists some $c > 1$ such that
$$K - 1 > \frac{4(c + 1)}{(\delta_2 - 1 - \alpha)}.$$
Set $s = (c + 1)(K - 1)^{-1}$. Then $4 s < \delta_2 - 1 - \alpha$ and
$$(1 + s) \frac{d - r_t}{d} > (1+s)\frac{K - 1}{K} = 1 + \frac{c}{K}.$$
Hence, for some $c > 1$,
\begin{equation*}
P[\cup_{\{d > K|t|\}} V(d)] \leq c' p_n^{-c}.
\end{equation*}
Restricting to the event $\cap_{\{d>K |t|\}}\{V(d)\}^c$, by Lemma \ref{lemma1} and Condition (G), we have
\begin{eqnarray*}
&&\sum_{k\in M_3} BF(k,t)\\ &\preceq& \sum_{k\in M_3}\left\{n\tau^2_{1n} \lambda_m (1-\phi_n)\right\}^{-(r_k-r_t)/2} \lambda_m^{-|t|/2}q^{|k|-|t|}_n p^{(1+4s)(r_k-r_t)}_n\\
&\preceq& \sum_{k\in M_3} \left(\frac{p_n^{1 + 4s}q_n}{p^{1+\delta_2}_n  \vee \sqrt{n}}\right)^{r_k-r_t}
\lambda_m^{-|t|/2}.
\end{eqnarray*}
Note that $r_k - r_t > (K - 1)|t|$ for $k \in M_3$. By Conditions (B) and (G),
\begin{eqnarray*}
\sum_{k\in M_3} BF(k,t) &\preceq& \sum_{k\in M_3} \left(\frac{p_n^{4s + \alpha}}{p^{1+\delta_2}_n  \vee \sqrt{n}}\right)^{r_k-r_t} p_n^{(K - 1) |t|} \\
&\preceq& \sum_{k\in M_3} \left(\frac{p_n^{1 + 4s + \alpha}}{p^{1+\delta_2}_n  \vee \sqrt{n}}\right)^{r_k-r_t}\\
&\leq& \left(1 + \frac{p_n^{1 + 4s + \alpha}}{p^{1+\delta_2}_n  \vee \sqrt{n}}\right)^{p_n} - 1.
\end{eqnarray*}
It is easy to show that
$$p_n \log\left(1 + \frac{p_n^{1 + 4s + \alpha}}{p^{1+\delta_2}_n  \vee \sqrt{n}}\right) \sim p_n\frac{p_n^{1 + 4s + \alpha}}{p^{1+\delta_2}_n \vee \sqrt{n}} \leq {p_n^{1 + 4s + \alpha - \delta}} = o(1).$$
Therefore, with probability at least $1 - c' p^{-c}$ for some $c > 1$,
\begin{equation}\label{eq8}
\sum_{k \in M_3} BF(k, t) {\rightarrow} 0.
\end{equation}

{\bf Under-fitted models.} Note that
\begin{eqnarray*}
R^*_k - R^*_{k \vee t} &=& Y' (P_{k\vee t} - P_k) Y \\
&=&\|(P_{{k\vee t}} - P_{k})Y\|^2_2 \\
&\geq& \{ \|(P_{{k\vee t}} - P_{k})Z_t\beta_t\|_2 - \|(P_{{k\vee t}} - P_{k})\epsilon\|_2\}^2.
\end{eqnarray*}
By definition, $\| (P_{{k\vee t}} - P_{k})Z_t\beta_t \|_2 = \|(I-P_{k})Z_t\beta_t\|_2 \geq \sqrt{\Delta}$ for all $k \in M_4$. We have, for any $w' \in (0,1)$,
\begin{eqnarray*}
&&P[ \cup_{k \in M_4} \{R^*_k - R^*_{k\vee t} < (1-w')^2 \Delta\}] \\
&\leq& P[ \cup_{k \in M_4}\{\|(P_{{k\vee t}} - P_{k})\epsilon\|_2 > w' \sqrt{\Delta} \} ]\\
&\leq& P(\|P_t \epsilon\|_2 > w' \sqrt{\Delta})\\
&\leq& P\left(\frac{\chi^2_{|t|}}{|t|} > 1 + \frac{w'^2 {\Delta}}{\sigma^2 |t|} - 1\right)\\
&\leq& \exp(-c' \Delta).
\end{eqnarray*}
This implies that for $w \in (0,1)$,
\begin{eqnarray*}
&&P\left[ \cup_{k \in M_4} \{R_k - R_{k\vee t} < \Delta(1-w)\} \right]\nonumber \\
&\leq& P\left[ \cup_{k \in M_4} \{R_k^* - R_{k\vee t} < \Delta(1-w)\} \right]\nonumber\\
&\leq& P\left[ \cup_{k \in M_4} \{R^*_k - R^*_{k\vee t} < \Delta(1-w/2)\} \right]\\
& & + P\left[ \cup_{k \in M_4} \{R_{k\vee t}-R^*_{k\vee t} > \Delta w/2\} \right].\nonumber\\
&\leq& \exp(-c' \Delta) + P\left[ \cup_{k \in M_4} \{R_{k\vee t}-R^*_{k\vee t} > \Delta w/2\} \right].\nonumber
\end{eqnarray*}
Let $Z_{k\vee t} = U_{n\times |k\vee t|} \Lambda_{|k\vee t|\times |k\vee t|} V'_{|k\vee t|\times |k\vee t|}$ be the SVD of $Z_{k\vee t}$. Then
\begin{eqnarray*}
R_{k\vee t}-R^*_{k\vee t} &=& Y' \{I - Z_{k \vee t}(\tau_{1n}^{-2}I + Z_{k \vee t}' Z_{k \vee t})^{-1}Z_{k \vee t}'\}Y  - Y'(I - P_{k \vee t})Y \\
&=& Y' \{P_{k \vee t}- Z_{k \vee t}(\tau_{1n}^{-2}I + Z_{k \vee t}' Z_{k \vee t})^{-1}Z_{k \vee t}'\}Y\\
&=& Y' \{UU'- U\Lambda V'(\tau_{1n}^{-2}VV' + V\Lambda^2 V')^{-1}V\Lambda U'\}Y\\
&=& Y'U \{I- \Lambda (\tau_{1n}^{-2}I +\Lambda^2)^{-1}\Lambda\} U'Y\\
&=& Y'U(I + \tau^{2}_{1n}\Lambda^2)^{-1}U'Y\\
&\leq& (s' n\tau^2_{1n} )^{-1} Y'UU'Y.
\end{eqnarray*}
Since $U_{n\times |k\vee t|}$ is an unitary matrix with rank at most $(K+1)|t|$, by Condition (D), we have
\begin{eqnarray*}
P(R_{k\vee t}-R^*_{k\vee t} > \Delta w/2) &=& P(Y'UU'Y > s' n\tau^2_{1n} \Delta w/2) \\
&\leq& P(\epsilon'P_{k \vee t}\epsilon > w' n\tau^2_{1n} \Delta) \\
&=& P\left(\frac{\chi^2_{|{k \vee t}|}}{|{k \vee t}|}  > 1 + \frac{w' n\tau^2_{1n} \Delta}{\sigma^2 |{k  \vee t}|} - 1\right) \\
&\preceq& \exp(-c' n \tau_{1n}^2\Delta),
\end{eqnarray*}
and hence,
\begin{eqnarray*}
P\left[ \cup_{k \in M_4} \{R_{k\vee t}-R^*_{k\vee t} > \Delta w/2\} \right] &\preceq& p_n^{K |t| + 1} \exp(-c' n \tau_{1n}^2 \Delta)\\
&\preceq& \exp(-s' \Delta).
\end{eqnarray*}
Therefore,
\begin{eqnarray*}\label{eq10}
P[ \cup_{k \in M_4} \{R_k - R_{k\vee t} < \Delta(1-w)\}] \leq 2\exp(-c' \Delta).
\end{eqnarray*}
This, together with Lemmas \ref{lemma2} and \ref{lemma3}, for $0 < c=3w < 1$, we have
\begin{eqnarray*}
& & P[ \cup_{k \in M_4} \{\tilde{R}_k - \tilde{R}_t < \Delta (1-c)\} ] \\
&\leq& P[ \cup_{k \in M_4} \{\tilde{R}_k - \tilde{R}_{k\vee t} < \Delta(1-2w)\} ] + P[ \cup_{k \in M_4} \{\tilde{R}_{k\vee t} - \tilde{R}_t < -\Delta w\} ]\\
&\leq&  P[ \cup_{k \in M_4} \{R_k - R_{k\vee t} < \Delta(1-w)\} ] + P[ \cup_{k \in M_4} \{R_t - R_{k\vee t} > w^2\Delta\}]\\
&\leq& 2\exp(-c' \Delta) + P[ \cup_{k \in M_4} \{R_t - R_{k\vee t}^* > w^2\Delta\}]\\
&\leq& 2\exp(-c' \Delta) + P[ \cup_{k \in M_4} \{R^*_t - R^*_{k\vee t} > w^2\Delta/2\}] + P( R_t - R^*_t > w^2\Delta/2)\\
&\leq& 3\exp(-w' \Delta) + P[ \cup_{k \in M_4} \{R^*_t - R^*_{k\vee t} > w^2\Delta/2\} ].
\end{eqnarray*}
It is easy to show that
\begin{eqnarray*}
P(R_t^* - R_{k\vee t}^* > w^2\Delta/2)
&=& P\left(\frac{\chi^2_{|{k \wedge t^c}|}}{|{k \wedge t^c}|} > 1 + \frac{w^2\Delta/2}{\sigma^2 |{k \wedge t^c}|} - 1\right) \\
&\preceq& \exp(-c' \Delta).
\end{eqnarray*}
By Condition (E),
\begin{eqnarray*}
P\left[ \cup_{k \in M_4} \{R^*_t - R^*_{k\vee t} > w^2\Delta / 2\} \right] &\preceq& p_n^{K |t| + 1} \exp(-c' \Delta)\\
&\preceq& \exp(-w' \Delta).
\end{eqnarray*}
We have, for some $c > 1$,
\begin{eqnarray*}
P[ \cup_{k \in M_4} \{\tilde{R}_k - \tilde{R}_t < \Delta (1-c)\}] \leq 4 \exp(-c' \Delta) \leq s' p_n^{-c}.
\end{eqnarray*}

Restricting to the event $\{\tilde{R}_k - \tilde{R}_t \geq \Delta (1-c), \forall k \in M_4\}$, by Condition (E), we have
\begin{eqnarray*}
&&\sum_{k\in M_4} BF(k,t)\\ &\preceq& \sum_{k\in M_4} \{n\tau^2_{1n} \lambda_m (1 - \phi_n)\}^{|t|/2} \lambda_m^{-|t|/2} q^{|k|-|t|}_n \exp\{-\Delta (1-c) / 2\sigma^2 \} \\
&\preceq& \sum_{k\in M_4} (n\tau^2_{1n})^{|t|/2} q^{|k|-|t|}_n \exp\{-\Delta (1-c) / 2\sigma^2 \} \\
&\preceq& \exp\left[-\frac{1}{2\sigma^2} \{\Delta(1-c) - \sigma^2|t|\log (p^{2+2\delta_1}_n \vee n) \} \right] \sum_{k\in M_4} q^{-|t|}_n \\
&\preceq& \exp\left[-\frac{1}{2\sigma^2} \{\Delta(1-c) - \sigma^2|t|\log (p^{2+2\delta_1}_n \vee n) - 2\sigma^2 {(K + 1 - \alpha)|t|}\log p_n\} \right] \\
&\rightarrow& 0.
\end{eqnarray*}
Therefore, with probability at least $1 - s' p_n^{-c}$ for some $c > 1$, we have
\begin{equation}\label{eq11}
\sum_{k \in M_4} BF(k, t) {\rightarrow} 0.
\end{equation}

Combining (\ref{eq2}), (\ref{eq7}), (\ref{eq8}), and (\ref{eq11}), the proof is complete.

{\it Proof of Theorem 2.} Note that
\begin{eqnarray*}
P(\bfgamma = \mathcal{G} \mid \X, \bfsigma^2) &\geq& P[\cap_{i = 1}^{p_n} \{\bfgamma^i = t^i\} \mid \X, \bfsigma^2] \\
&=& 1 - P[\cup_{i = 1}^{p_n} \{\bfgamma^i = t^i\}^c \mid  \X, \bfsigma^2] \\
&\geq& 1 - \sum_{i = 1}^{p_n} P[\{\bfgamma^i = t^i\}^c \mid \X, \sigma_i^2].
\end{eqnarray*}
Under (D')-(G'), $c, s$ and $r_n$ in Theorem 1 can be chosen to be independent of the node index. Hence, as $n \rightarrow \infty$,
\begin{eqnarray*}
P(\bfgamma = \mathcal{G} \mid \X, \bfsigma^2) \geq 1 - c \sum_{i = 1}^{p_n} \frac{1}{p_n^{s}} \geq 1 - c \frac{1}{p_n^{s - 1}} \rightarrow 1.
\end{eqnarray*}
The proof is complete.

\bibliographystyle{agsm} 
\bibliography{refsup}